\documentclass[preprint,aps,12pt,showpacs,nofootinbib,tightenlines]{revtex4}
\usepackage{amsmath}
\usepackage{amssymb}
\usepackage{epsfig}
\usepackage{graphicx}
\begin{document}
%\preprint{NJNU-TH-07-05}
%%%%%%%%%%%%%%%%%%%%%%%%%%%%%%%%%%%%%%%%%%%%%
%%%%%%%%%%%%%%%%%%%%%%%%%%%%%%%%%%%%%%%%%%%%%%%%%%%%%%
\def\pslash{\rlap{\hspace{0.02cm}/}{p}}
\def\eslash{\rlap{\hspace{0.02cm}/}{e}}
%%%%%%%%%%%%%%%%%%%%%%%%%%%%%%%%%%%%%%%%%%%%%%%%%%%%%%
\title {The Study of the contribution of the LHT model to $Zb\bar{b}$ coupling }
\author{ Bingfang Yang$^1$$^,$$^2$}
\author{ Xuelei Wang$^1$}
\author{ Jinzhong Han$^1$}
\affiliation{ $^1$ College of Physics and Information Engineering,
Henan Normal University, Xinxiang 453007, China \\
$^2$ Basic Teaching Department, Jiaozuo University, Jiaozuo 454000,
China
     \vspace*{1.5cm}}
%\date{\today}
\begin{abstract}
In the framework of the Littlest Higgs Model with T-parity (LHT), we
study the contributions of the new particles to $Zb\bar{b}$
couplings at one-loop level. Based on these results, we further
study the branching ratio $R_{b}$ and the unpolarized
forward-backward asymmetry ${A_{FB}^{b}}$. We find that the
correction of the new particles to $Zb\bar{b}$ couplings is mainly
on the left-handed coupling and has small part of the parameter
space to alleviate the deviation between theoretical predictions and
experimental values. The precision measurement value of $R_{b}$ can
give severe constraints on the relevant parameters. The constraints
from the precision measurement value of ${A_{FB}^{b}}$ are very
weak.
\end{abstract}
 \maketitle
%%%%%%%%%%%%%%%%%%%%%%%%%%%%%%%%%%%%%%%%%%%%%%%%%%%%%%%%%%%%%%%%%%%%%%%%%%%%%%%
\section{ Introduction}
\noindent The Standard Model (SM) has been very successful, however,
it is still believed to be a theory effective at the electroweak
scale and some new physics (NP) must exist at higher energy regimes.
So far there have been many speculations on the possible forms of
the NP beyond the SM, one of the interesting possibilities is the
Little Higgs model. The little Higgs theory was proposed \cite{1} as
a possible solution to the hierarchy problem and remains a popular
candidate for the NP. The Littlest Higgs (LH) model \cite{2} is a
cute economical implementation of the little Higgs, but suffered
from severe constraints from electroweak precision tests \cite{3},
which would require raising the mass scale of the new particles to
far above TeV scale and thus reintroduce the fine-tuning in the
Higgs potential \cite{4}. The most serious constraints resulted from
the tree-level corrections to precision electroweak observables due
to the exchanges of the additional heavy gauge bosons present in the
theories, as well as from the small but non-vanishing vacuum
expectation value (VEV) of an additional weak-triplet scalar field.
In order to solve this problem, a discrete symmetry called T-parity
is proposed \cite{5}, which explicitly forbids any tree-level
contributions from the heavy gauge bosons to the observables
involving only the SM particles as external states. The interactions
that induce triplet VEV contributions is also forbidden. This model
is called the Littlest Higgs Model with T-parity (LHT). In the LHT
model, corrections to the precision electroweak observables are
generated exclusively at loop level.

The branching ratio $R_{b}$ is very sensitive to the NP beyond the
SM, the precision experimental value of $R_{b}$ may give a severe
constraint on the NP \cite{6}. Experimentally, the electroweak
observables have been precisely measured at the SLC and LEP, in the
most recent analysis of the electroweak data,
$R_{b}=0.21629\pm0.00066$ differs from the SM fit by $0.7\sigma$,
$A_{FB}^{b}=0.0992\pm0.0016$ disagrees with the SM fit by
$-2.9\sigma$ \cite{7}. Furthermore, the experimental value of
$Zb\bar{b}$ couplings disagrees with the SM fit by about $3\sigma$,
especially the deviation of the right-handed coupling is so large
that it is very difficult to explain. These significant deviations
from the ${A_{FB}^{b}}$ and the $Zb\bar{b}$ couplings might be the
first window into the NP. With the running of the LHC, they will be
further researched. In the LHT model, there are new fermions and new
gauge bosons, which can contribute to the $Zb\bar{b}$ couplings and
give modifications to the $R_{b}$ and $A_{FB}^{b}$. Therefore, it is
possible to give some constraints on the relevant parameters via
their radiative corrections to the $R_{b}$ and $A_{FB}^{b}$. In this
paper, we calculate the contributions of the LHT model to the
$Zb\bar{b}$ couplings. On this basis, we further study the $R_{b}$
and ${A_{FB}^{b}}$, then we give the constraints on the relevant
parameters according to the precision measurements.

This paper is organized as follows. In Sec.II we recapitulate the
LHT model and discuss the new flavor interactions which will
contribute to the $Zb\bar{b}$ vertex. In Sec.III we calculate the
one-loop contributions of the LHT model to the $Zb\bar{b}$ vertex,
$R_{b}$ and $A_{FB}^{b}$, then the relevant numerical results are
shown. Finally, we give our conclusions in Sec.IV.

\section{ A brief review of the LHT model}
 \noindent The LHT \cite{5} is based on a non-linear sigma model describing the
 spontaneous breaking of a global $SU(5)$ down to a global $SO(5)$. This
 symmetry breaking takes place at the scale $f\sim\mathcal{O}(TeV)$ and originates from the
 VEV of an $SU(5)$ symmetric tensor $\Sigma $, given by
 \begin {equation}
\Sigma_{0}\equiv<\Sigma>=
\begin{pmatrix}
0_{2\times2}&0&1_{2\times2}\\
0&1&0\\
1_{2\times2}&0&0_{2\times2}
\end{pmatrix}
\end{equation}
From the $SU(5)/SO(5)$  breaking, there arise 14 Goldstone bosons
which are described by the ``pion" matrix $\Pi$, given explicitly by
\begin {equation}
\Pi=
\begin{pmatrix}
-\frac{\omega^0}{2}-\frac{\eta}{\sqrt{20}}&-\frac{\omega^+}{\sqrt{2}}
&-i\frac{\pi^+}{\sqrt{2}}&-i\phi^{++}&-i\frac{\phi^+}{\sqrt{2}}\\
-\frac{\omega^-}{\sqrt{2}}&\frac{\omega^0}{2}-\frac{\eta}{\sqrt{20}}
&\frac{v+h+i\pi^0}{2}&-i\frac{\phi^+}{\sqrt{2}}&\frac{-i\phi^0+\phi^P}{\sqrt{2}}\\
i\frac{\pi^-}{\sqrt{2}}&\frac{v+h-i\pi^0}{2}&\sqrt{4/5}\eta&-i\frac{\pi^+}{\sqrt{2}}&
\frac{v+h+i\pi^0}{2}\\
i\phi^{--}&i\frac{\phi^-}{\sqrt{2}}&i\frac{\pi^-}{\sqrt{2}}&
-\frac{\omega^0}{2}-\frac{\eta}{\sqrt{20}}&-\frac{\omega^-}{\sqrt{2}}\\
i\frac{\phi^-}{\sqrt{2}}&\frac{i\phi^0+\phi^P}{\sqrt{2}}&\frac{v+h-i\pi^0}{2}&-\frac{\omega^+}{\sqrt{2}}&
\frac{\omega^0}{2}-\frac{\eta}{\sqrt{20}}
\end{pmatrix}
\end{equation}
Under T-parity the SM Higgs doublet,
$H=(-i\pi^+\sqrt{2},(v+h+i\pi^0)/2)^T$ is T-even while other fields
are T-odd.

The Goldstone bosons $\omega^{\pm},\omega^{0},\eta$ are respectively
eaten by the new T-odd gauge bosons $W_{H}^{\pm},Z_{H},A_{H}$, which
obtain masses at $\mathcal O(\upsilon^{2}/f^{2})$
\begin {equation}
M_{W_{H}}=M_{Z_{H}}=gf(1-\frac{\upsilon^{2}}{8f^{2}}),M_{A_{H}}=\frac{g'f}{\sqrt{5}}
(1-\frac{5\upsilon^{2}}{8f^{2}})
\end {equation}
with $g$ and $g'$ being the SM $SU(2)$ and $U(1)$ gauge couplings,
respectively.

The Goldstone bosons $\pi^{\pm},\pi^{0}$ are eaten by the T-even
$W_{L}^{\pm}$and $Z_{L}$ bosons of the SM, which obtain masses at
$\mathcal O(\upsilon^{2}/f^{2})$
\begin {equation}
M_{W_{L}}=\frac{g\upsilon}{2}(1-\frac{\upsilon^{2}}{12f^{2}}),M_{Z_{L}}=\frac{g\upsilon}
{2\cos\theta_{W}}(1-\frac{\upsilon^{2}}{12f^{2}})
\end {equation}
The photon $A_{L}$ is also T-even and remains massless.

For each SM fermion, a copy of mirror fermion with T-odd quantum
number is added in order to preserve the T-parity. For the mirror
quarks, we denote them by $u_{H}^{i},d_{H}^{i}$, where i= 1, 2, 3
are the generation index. At the order of $\mathcal
O(\upsilon^{2}/f^{2})$ their masses are given by
\begin{equation}
m_{d_{H}^{i}}=\sqrt{2}\kappa_if, m_{u_{H}^{i}}=
m_{d_{H}^{i}}(1-\frac{\upsilon^2}{8f^2})
\end{equation}
where $\kappa_i$ are the diagonalized Yukawa couplings of the mirror
quarks.

In order to cancel the quadratic divergence of the Higgs mass
induced by top loops, an additional heavy quark $T^{+}$ is
introduced, which is even under T-parity. The implementation of
T-parity then requires also a T-odd partner $T^{-}$. Their masses
are given by
\begin{eqnarray}
m_{T^{+}}&=&\frac{f}{v}\frac{m_{t}}{\sqrt{x_{L}(1-x_{L})}}[1+\frac{v^{2}}{f^{2}}(\frac{1}{3}-x_{L}(1-x_{L}))]\\
m_{T^{-}}&=&\frac{f}{v}\frac{m_{t}}{\sqrt{x_{L}}}[1+\frac{v^{2}}{f^{2}}(\frac{1}{3}-\frac{1}{2}x_{L}(1-x_{L}))]
\end{eqnarray}
where $x_{L}$ is the mixing parameter between the SM top-quark $t$
and the new top-quark $T^{+}$.

Just like the SM, the mirror sector in the LHT model also has weak
mixing, parameterised by unitary mixing matrices: two for mirror
quarks and two for mirror leptons:
\begin{equation}
V_{Hu},V_{Hd},V_{Hl},V_{H\nu}
\end{equation}
 $V_{Hu}$ and $V_{Hd}$ are for the mirror quarks which are present in our
analysis. $V_{Hu}$ and $V_{Hd}$ satisfy the physical constraints
$V_{Hu}^{\dag}V_{Hd}=V_{CKM}$. We follow \cite{8} to parameterize
$V_{Hd}$ with three angles
$\theta^d_{12},\theta^d_{23},\theta^d_{13}$ and three phases
$\delta^d_{12},\delta^d_{23},\delta^d_{13}$
\begin{eqnarray}
V_{Hd}=
\begin{pmatrix}
c^d_{12}c^d_{13}&s^d_{12}c^d_{13}e^{-i\delta^d_{12}}&s^d_{13}e^{-i\delta^d_{13}}\\
-s^d_{12}c^d_{23}e^{i\delta^d_{12}}-c^d_{12}s^d_{23}s^d_{13}e^{i(\delta^d_{13}-\delta^d_{23})}&
c^d_{12}c^d_{23}-s^d_{12}s^d_{23}s^d_{13}e^{i(\delta^d_{13}-\delta^d_{12}-\delta^d_{23})}&
s^d_{23}c^d_{13}e^{-i\delta^d_{23}}\\
s^d_{12}s^d_{23}e^{i(\delta^d_{12}+\delta^d_{23})}-c^d_{12}c^d_{23}s^d_{13}e^{i\delta^d_{13}}&
-c^d_{12}s^d_{23}e^{i\delta^d_{23}}-s^d_{12}c^d_{23}s^d_{13}e^{i(\delta^d_{13}-\delta^d_{12})}&
c^d_{23}c^d_{13}
\end{pmatrix}
\end{eqnarray}
\section{The one-loop corrections to $Zb\bar{b}$ couplings in the LHT model}
\noindent We employ the following notation for the effective
$Zb\bar{b}$ interaction:
\begin{eqnarray}
L_{Zb\bar{b}}&=&\frac{e}{S_{W}C_{W}}(g_{L}^{b}\bar{b}\gamma^{\mu}bP_{L}+g_{R}^{b}\bar{b}\gamma^{\mu}bP_{R})Z_{\mu} \nonumber\\
&=&\frac{e}{2S_{W}C_{W}}\bar{b}\gamma^{\mu}(g_{V}^{b}-g_{A}^{b}\gamma_{5}bZ_{\mu})
\end{eqnarray}
where $\theta_{W}$ is the Weinberg angle, $S_{W}=\sin\theta_{W}$,
$C_{W}=\cos\theta_{W}$, $P_{L}=\frac{1-\gamma_{5}}{2}$ and
$P_{R}=\frac{1+\gamma_{5}}{2}$. The effective couplings are then
written as
\begin{eqnarray}
\bar{g}_{L,R}^{b}&=&g_{L,R}^{b}+\delta g_{L,R}^{SM}+\delta
g_{L,R}^{NP}\\ \bar{g}_{V,A}^{b}&=&g_{V,A}^{b}+\delta
g_{V,A}^{SM}+\delta g_{V,A}^{NP}
\end{eqnarray}
where $\bar{g}_{L,R}^{b}, \bar{g}_{V,A}^{b}$ are respectively the
radiatively-corrected effective couplings, $g_{L,R}^{b}$ are
respectively the left-handed and right-handed $Zb\bar{b}$ couplings
at tree level, $\delta g_{L,R}^{SM} $ and $\delta g_{L,R}^{NP}$ are
their corresponding one-loop corrections of the SM and the NP,
$g_{V,A}^{b}$ are respectively the vector and axial vector coupling
coefficients of $Zb\bar{b}$ interaction at tree level, $\delta
g_{V,A}^{SM}$ and $\delta g_{V,A}^{NP}$ are their corresponding
one-loop corrections of the SM and the NP. The tree-level couplings
are given by
\begin{eqnarray}
g_{L}^{b}&=&-\frac{1}{2}+\frac{1}{3}S_{W}^{2}~~~,~~~g_{R}^{b}=\frac{1}{3}S_{W}^{2}\\
g_{V}^{b}=g_{L}^{b}+g_{R}^{b}&=&-\frac{1}{2}+\frac{2}{3}S_{W}^{2}
~~~,~~~g_{A}^{b}=g_{L}^{b}-g_{R}^{b}=-\frac{1}{2}
\end{eqnarray}

The branching ratio is defined as
\begin{equation}
R_{b}=\frac{\Gamma(Z\rightarrow b\bar{b})}{\Gamma(Z\rightarrow
hadrons)}
\end{equation}

The full hadron width is the sum of widths of five quark channels:
\begin{equation}
\Gamma(Z\rightarrow hadrons)=\Gamma(Z\rightarrow
u\bar{u})+\Gamma(Z\rightarrow d\bar{d})+\Gamma(Z\rightarrow
s\bar{s})+\Gamma(Z\rightarrow c\bar{c})+\Gamma(Z\rightarrow
b\bar{b})
\end{equation}

For the decays to any of the five pairs of quarks $q\bar{q}$ we
have\cite{9}
\begin{equation}
\Gamma_{q}\equiv\Gamma(Z\rightarrow
q\bar{q})=12\Gamma_{0}(g_{Aq}^{2}R_{Aq}+g_{Vq}^{2}R_{Vq})
\end{equation}
with $\Gamma_{0}=\frac{G_{F}M_{Z_{L}}^{3}}{24\sqrt{2}\pi}$, here
$g_{Aq}$ and $g_{Vq}$ are the axial-vector and effective vector
couplings. The radiators  $R_{Aq}$ and $R_{Vq}$  contain
contributions from the final state gluons and photons. In the
crudest approximation
\begin{equation}
R_{Vq}=R_{Aq}=1+\frac{\hat{\alpha_{s}}}{\pi}
\end{equation}
where $\alpha_{s}(q^{2})$ is the QCD running coupling constant:
\begin{equation}
\hat{\alpha_{s}}\equiv\alpha_{s}(q^{2}=M_{Z_{L}}^{2})
\end{equation}

The expression of the radiative correction to $R_{b}$ can be
expressed as \cite{10}
\begin{equation}
\delta
R_{b}\simeq\frac{2R_{b}^{SM}(1-R_{b}^{SM})}{g_{Vb}^{2}(3-\beta^{2})+2g_{Ab}^{2}\beta^{2}}[g_{Vb}(3-\beta^{2})\delta
g_{Vb}+2g_{Ab}\beta^{2}\delta g_{Ab}]
\end{equation}
with $\beta=\sqrt{1-\frac{4\hat{m}_{b}^{2}}{M_{Z_{L}}^{2}}}$ being
the velocity of b-quark in $Z$ decay, here $\hat{m}_{b}$ is the
value of the running mass of the b-quark at scale $M_{Z_{L}}$
calculated in $\overline{MS}$ scheme \cite{11}.

The unpolarized forward-backward asymmetry in the decay to
$b\bar{b}$ equals:
\begin{equation}
A_{FB}^{b}=\frac{N_{F}-N_{B}}{N_{F}+N_{B}}
\end{equation}
 where $N_{F}$ is the cross section for finding the scattered
fermion in the hemisphere defined by the incident electron direction
and $N_{B}$ is the cross section for finding it in the positron
hemisphere.
 It can be expressed as
\begin{equation}
A_{FB}^{b}=\frac{3}{4}(1-\frac{k_{A}}{\pi})A_{e}A_{b}
\end{equation}
where the factor $(1-\frac{k_{A}}{\pi})$ represents a QCD radiative
correction, as in Ref. \cite{12}, for which we use the numerical
value 0.95, $A_{e}$ refers to the creation of $Z$ boson in
$e^{+}e^{-}$ -annihilation, while $A_{b}$ is the left-right coupling
constant asymmetry£¬refers to its decay in $b\bar{b}$ \cite{9}
\begin{equation}
A_{b}=\frac{2g_{Ab}g_{Vb}}{\beta^{2}g_{Ab}^{2}+(3-\beta^{2})g_{Vb}^{2}/2}
\end{equation}

 The relevant Feynman diagrams for the LHT contributions are
shown in Fig.1. We use the 't Hooft-Feynman gauge, so the
contributions of Goldstone bosons should be involved. In our
calculation, $g_{Ab}$ and $g_{Vb}$ should be replaced by
$\bar{g}_{Ab}$ and $\bar{g}_{Vb}$, $g_{V,A}^{b}+\delta g_{V,A}^{SM}$
can be found in Ref. \cite{13}. The calculations of the loop
diagrams are straightforward. Each loop diagram is composed of some
scalar loop functions \cite{14}, which are calculated by using
LOOPTOOLS \cite{15}. The relevant Feynman rules can be found in Ref.
\cite{16}. We applied the on-shell renormalization scheme and have
checked that the divergences are canceled.
\begin{figure}
\scalebox{0.38}{\epsfig{file=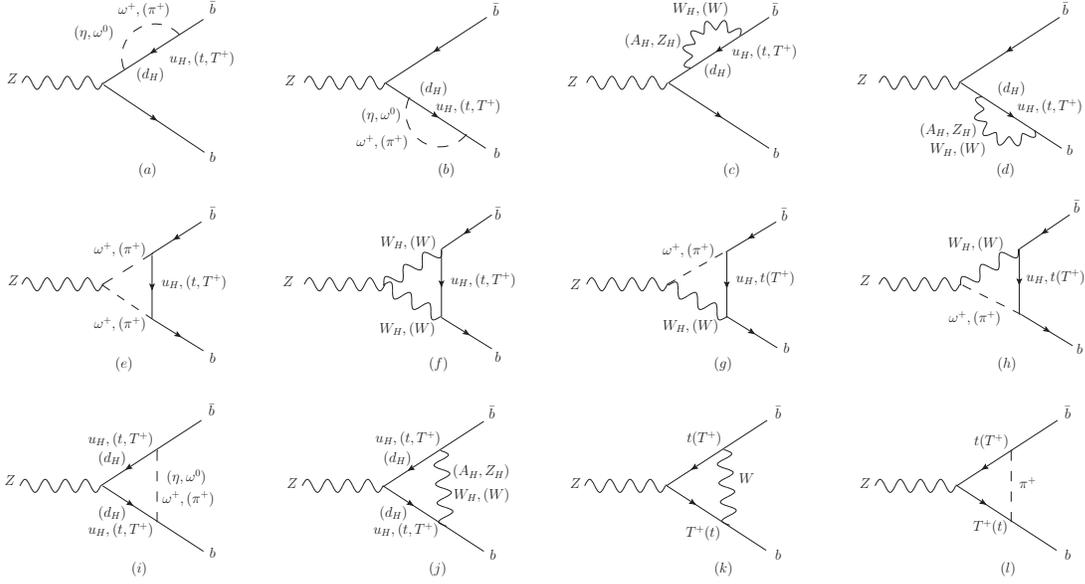}}
 \caption{Feynman diagrams of
$Z\rightarrow b\bar{b}$ at one-loop level in the LHT model.}
\end{figure}

In the numerical calculations we take the input parameters \cite{17}
as Fermi constant $G_{F}=1.16637\times 10^{-5}GeV^{-2}$, the
fine-structure constant $\alpha=1/128$, $Z$-boson mass
$M_{Z_{L}}=91.2GeV$, fermion masses $m_{f}$, the electroweak mixing
angle $S_{W}^{2}=0.231$ and the final-state asymmetry parameter
$A_{e}=0.1515$. In our calculation, the relevant LHT parameters are
the scale $f$, the mixing parameter $x_{L}$, the mirror quark masses
and parameters in the matrices $V_{Hu}$  and $V_{Hd}$.

For the mirror quark masses, from Eq.(5) we get
$m_{u_{H}^{i}}=m_{d_{H}^{i}}$ at $\mathcal O(\upsilon/f)$ and
further assume
\begin{equation}
m_{u_{H}^{1}}=m_{u_{H}^{2}}=m_{d_{H}^{1}}=m_{d_{H}^{2}}=M_{12},m_{u_{H}^{3}}=m_{d_{H}^{3}}=M_{3}
\end{equation}

For the matrices $V_{Hu}$ and $V_{Hd}$, considering the constraints
in Ref.\cite{18}, we study the completely generic scenario, i.e.the
six parameters of $V_{Hd}$ are arbitrary. After that, we follow
Ref.\cite{19} to consider the following two scenarios for
comparison:

Scenario I$:V_{Hd}=1,V_{Hu}=V_{CKM}^{\dag}$

Scenario II$
:S_{13}^{d}=0.5,\delta_{12}^{d}=\delta_{23}^{d}=0,\delta_{13}^{d}=\delta_{13}^{SM},S_{ij}^{d}=S_{ij}^{SM}$otherwise

\begin{figure}
\scalebox{0.61}{\epsfig{file=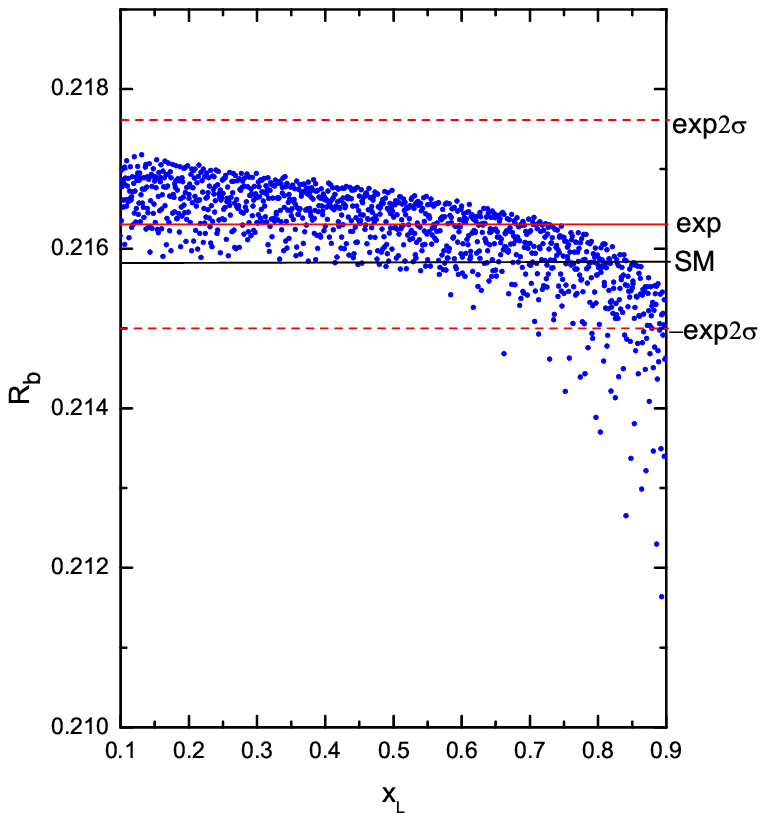}}
\scalebox{0.61}{\epsfig{file=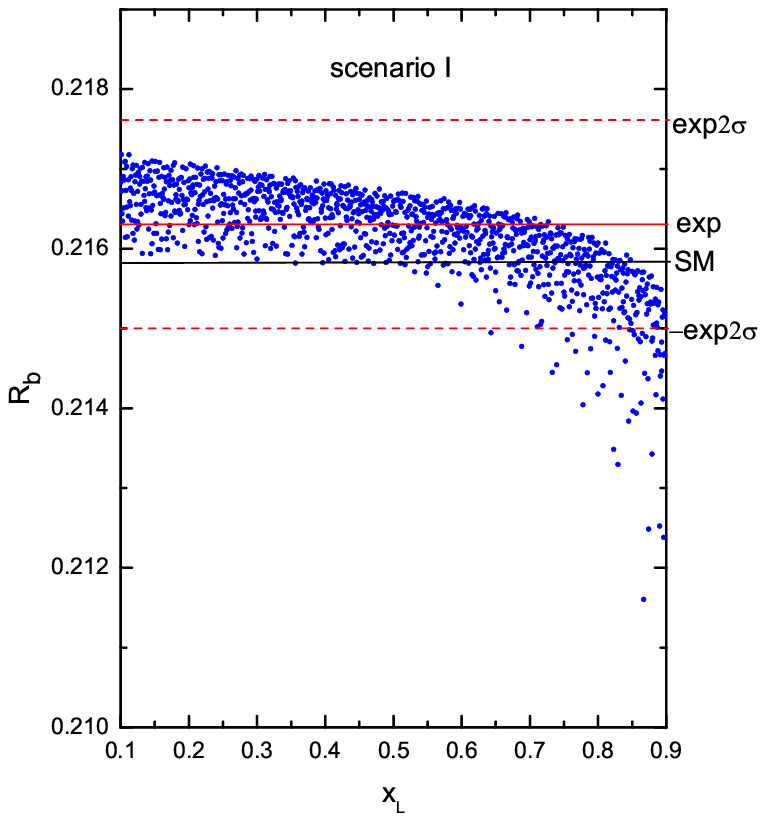}}
\scalebox{0.61}{\epsfig{file=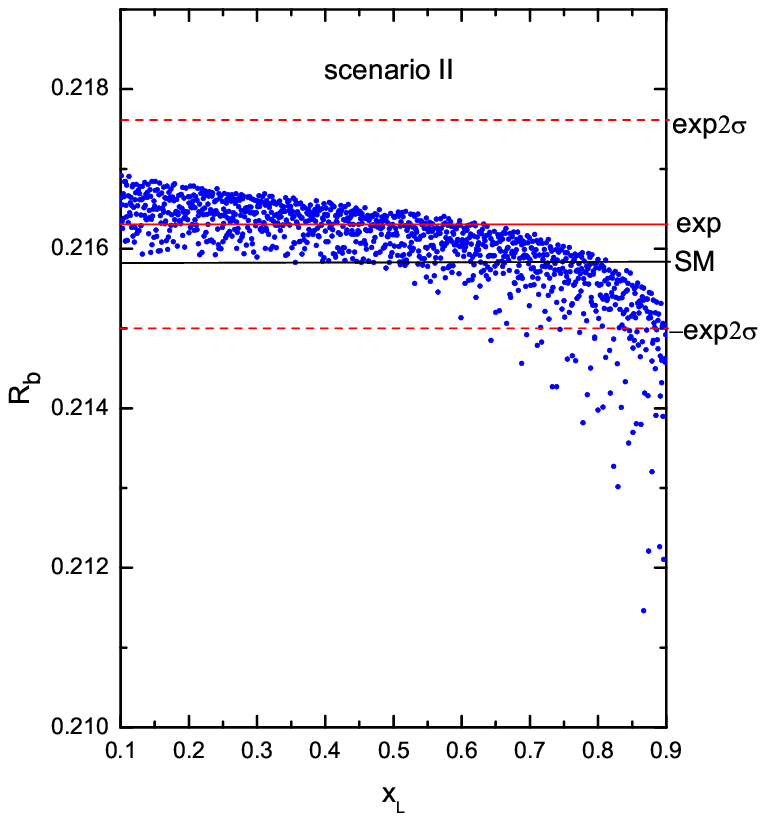}}\caption{Scatter plots of
$R_{b}$ versus $x_{L}$in arbitrary scenario, scenario I and scenario
II, respectively. The experimental value $R_{b}=0.21629\pm0.00066$.}
\end{figure}
\begin{figure}
\scalebox{0.61}{\epsfig{file=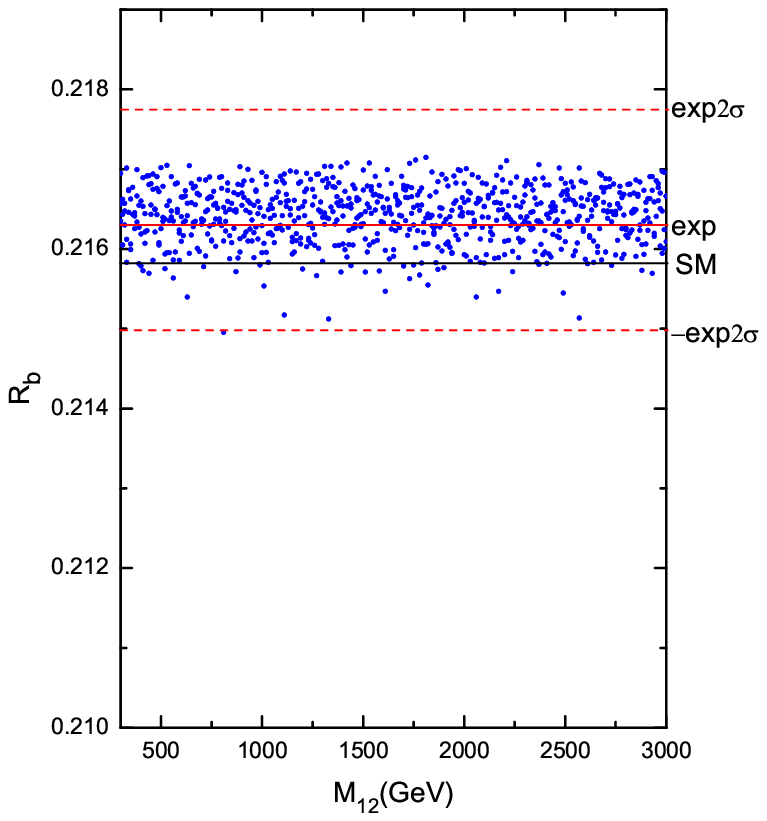}}
\scalebox{0.61}{\epsfig{file=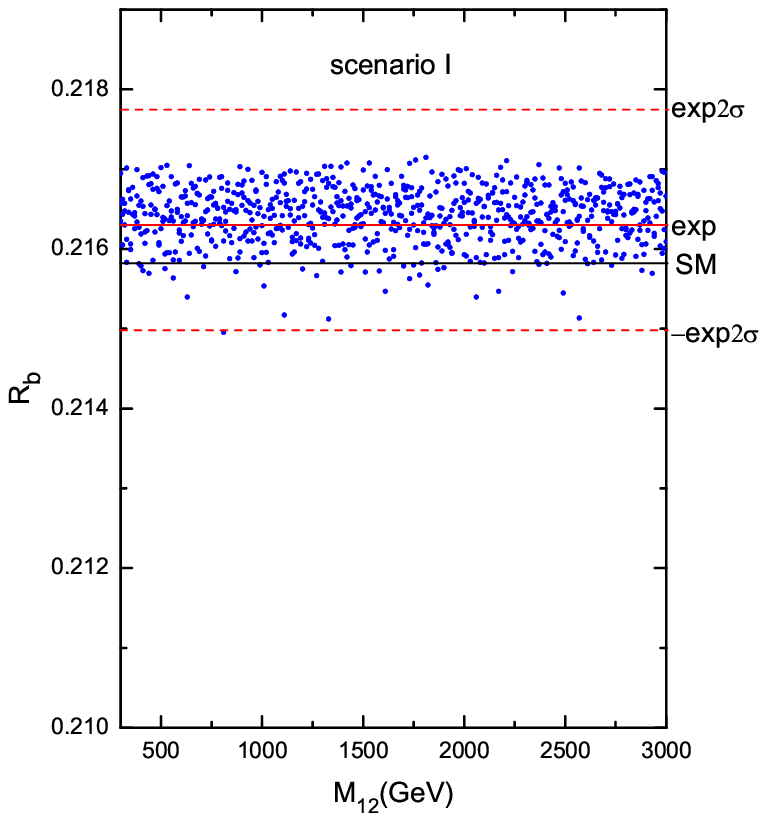}}
\scalebox{0.61}{\epsfig{file=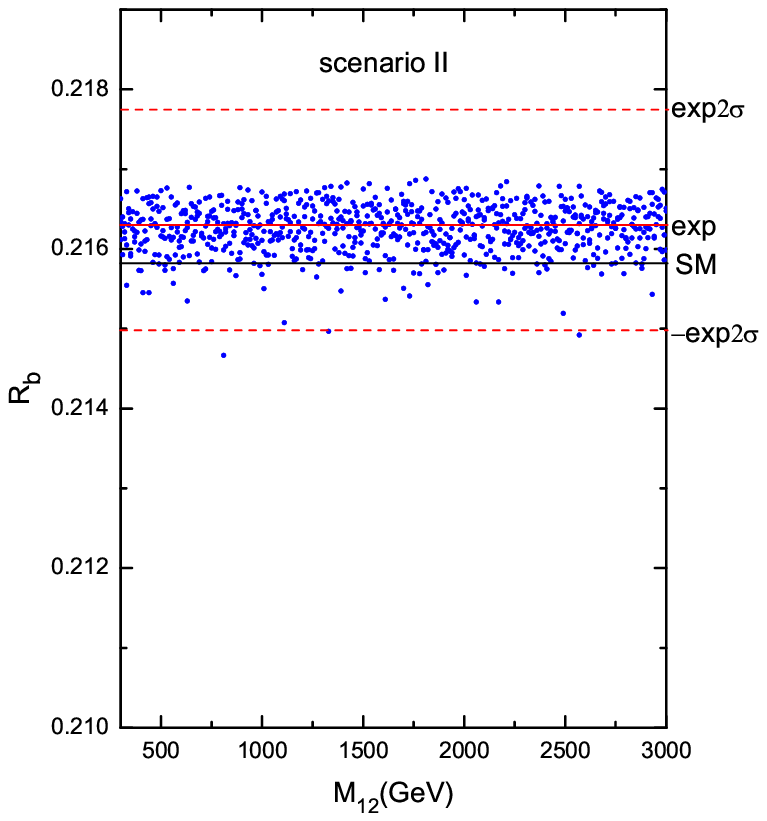}}\caption{Scatter plots of
$R_{b}$ versus $M_{12}$ in three different scenarios, respectively.
The experimental value $R_{b}=0.21629\pm0.00066$.}
\end{figure}
\begin{figure}
\scalebox{0.6}{\epsfig{file=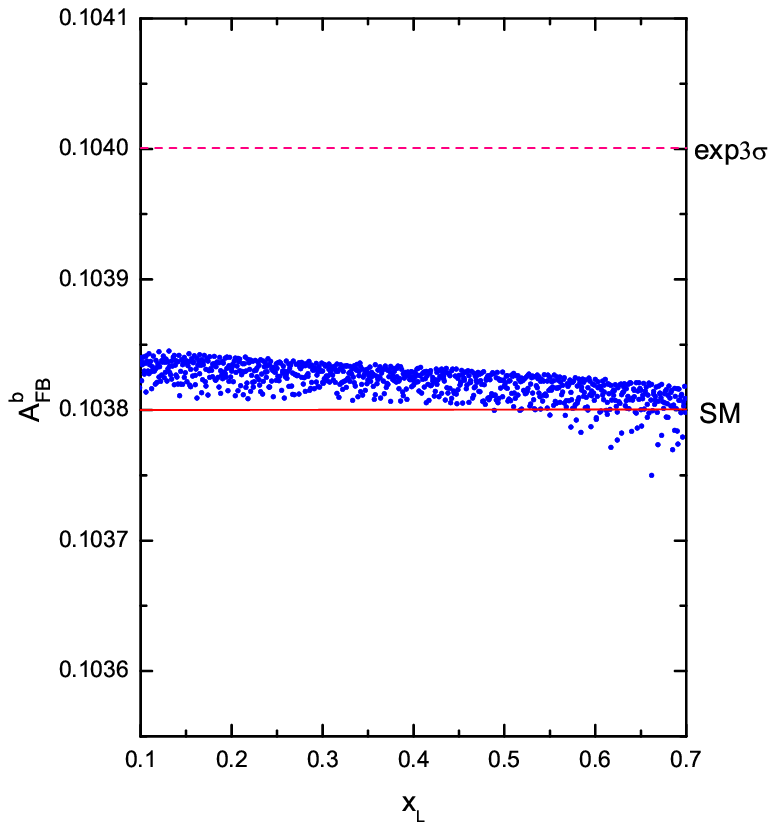}}
\scalebox{0.6}{\epsfig{file=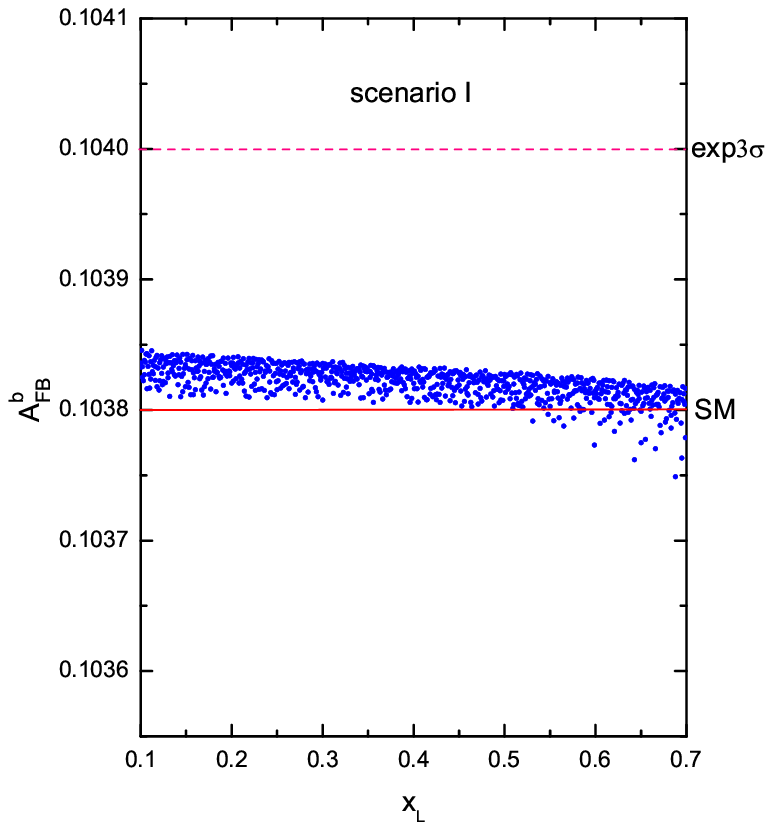}}
\scalebox{0.6}{\epsfig{file=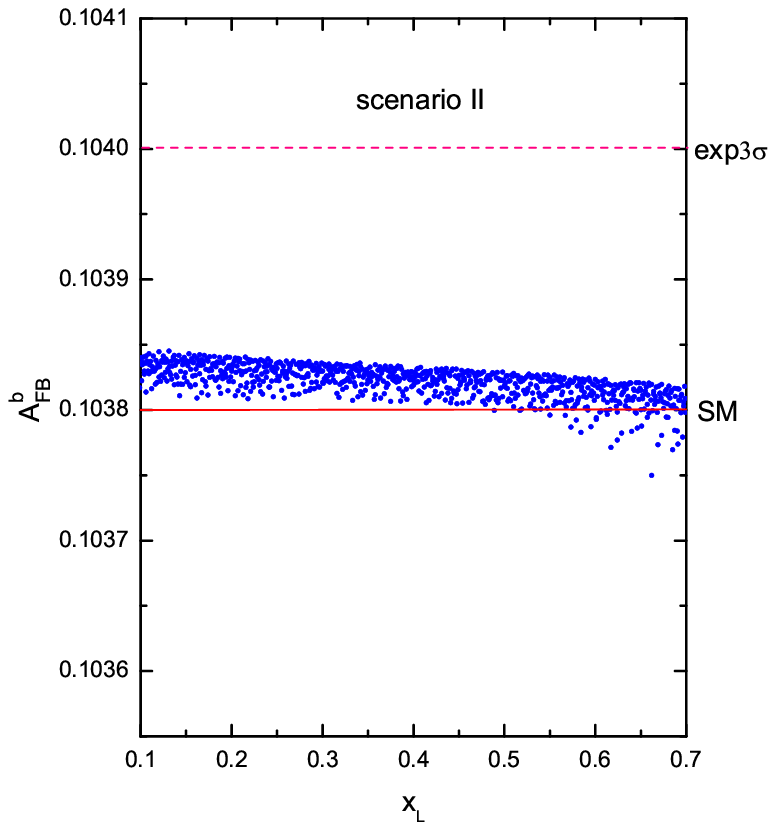}}\caption{Scatter plots of
$A_{FB}^{b}$ versus $x_{L}$ in three different scenarios,
respectively. The experimental value $A_{FB}^{b}=0.0992\pm0.0016$.}
\end{figure}
\begin{figure}
\scalebox{0.6}{\epsfig{file=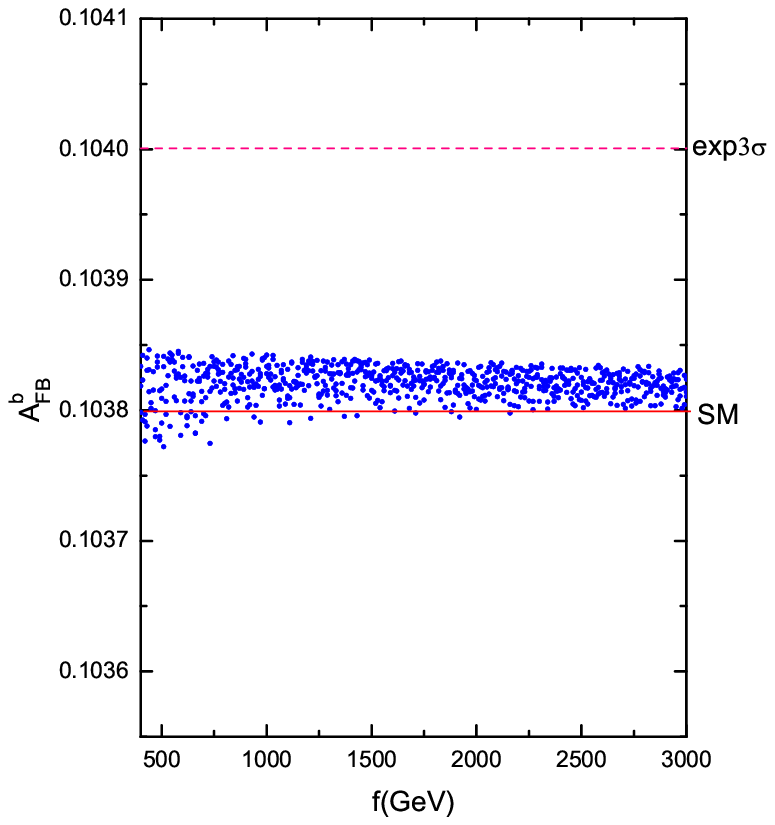}}
\scalebox{0.6}{\epsfig{file=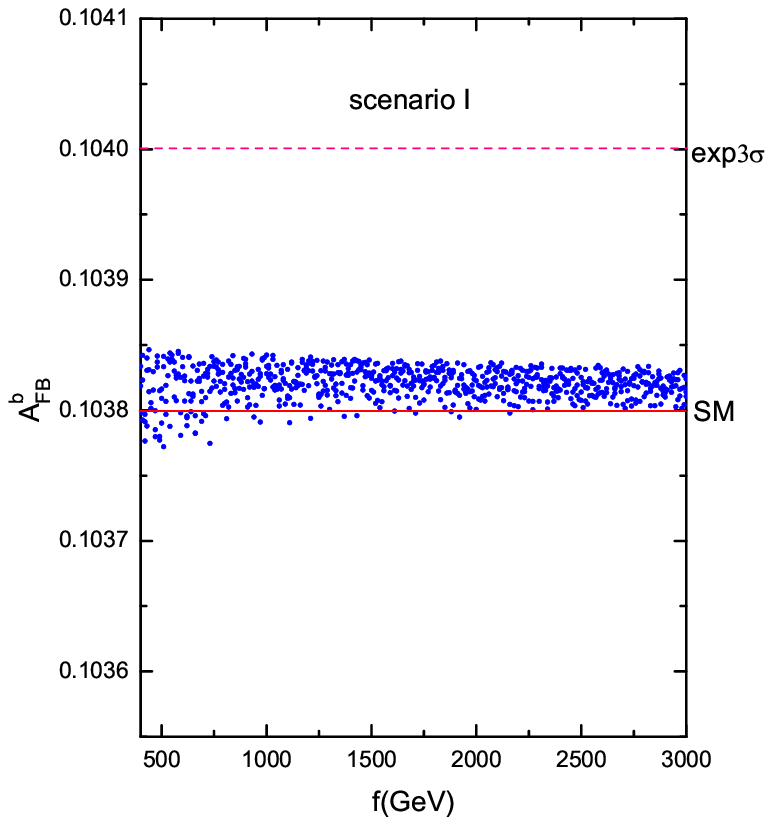}}
\scalebox{0.6}{\epsfig{file=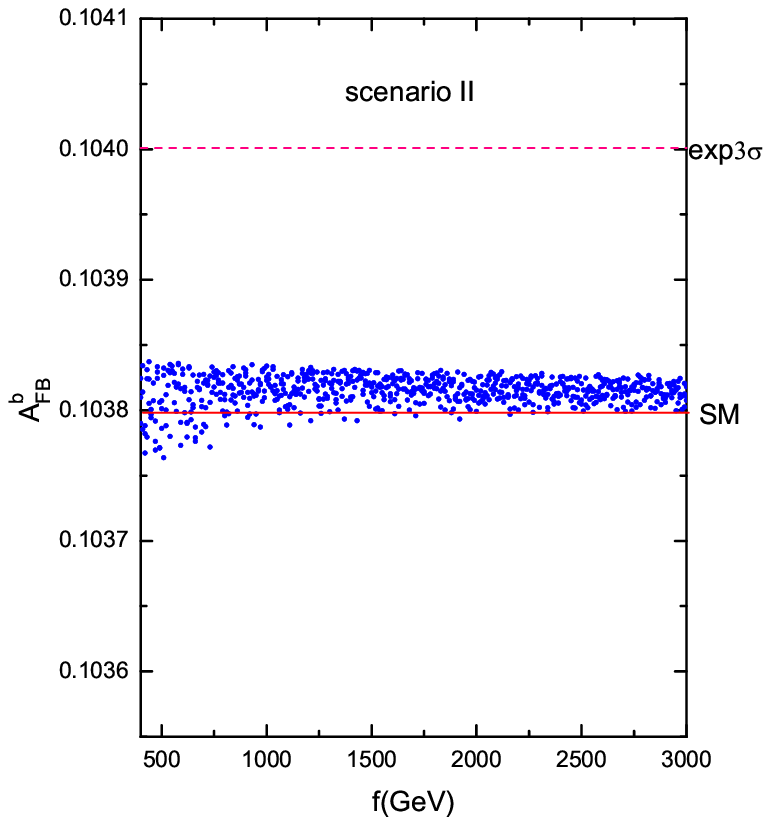}}
 \caption{Scatter plots of $A_{FB}^{b}$ versus $f$ in three different scenarios, respectively. The experimental value
$A_{FB}^{b}=0.0992\pm0.0016$.}
\end{figure}
\begin{figure}
\scalebox{0.6}{\epsfig{file=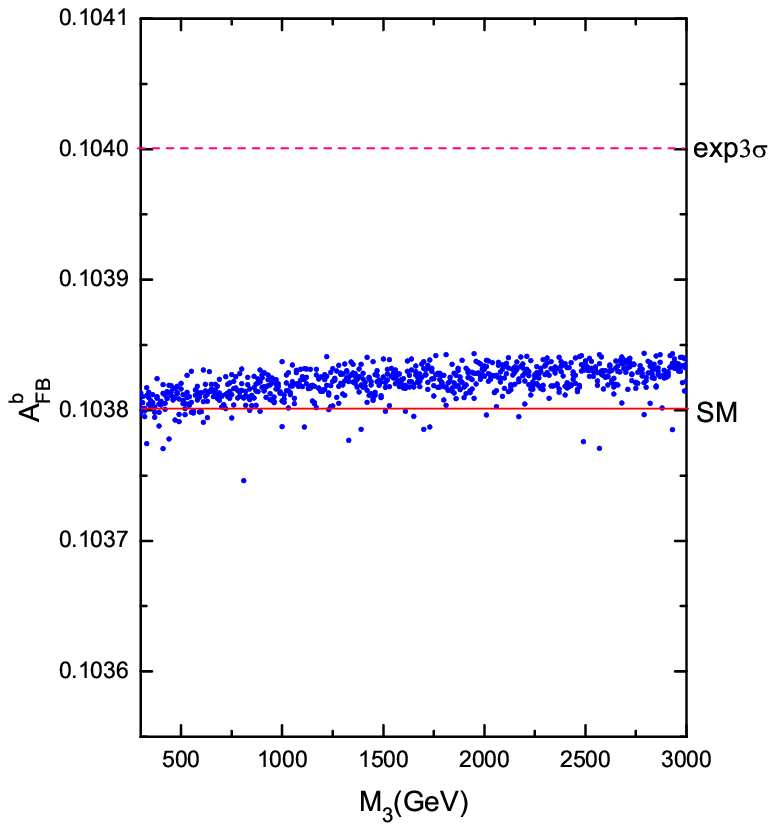}}
\scalebox{0.6}{\epsfig{file=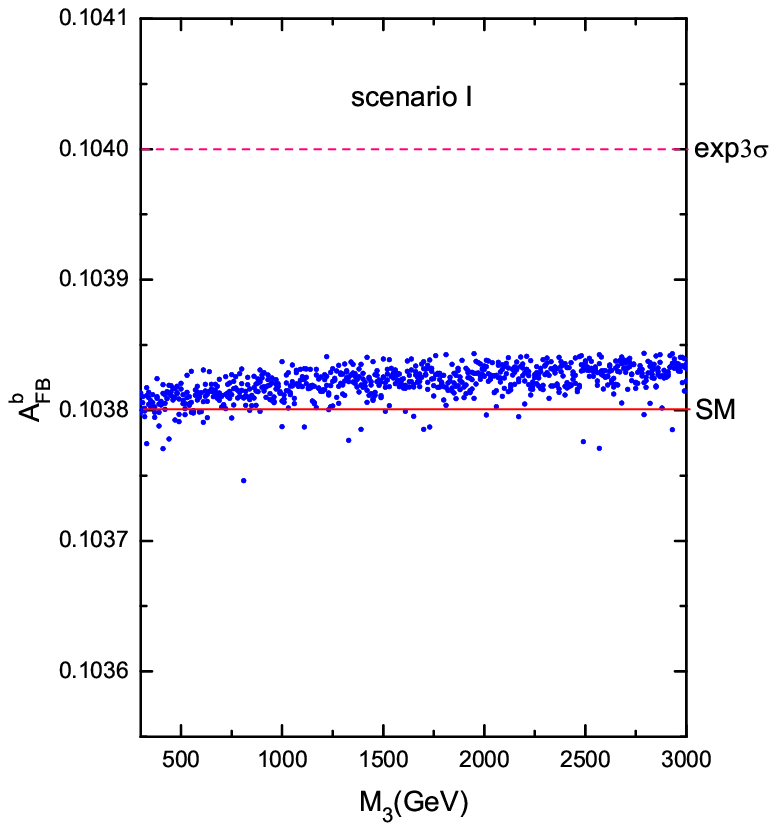}}
\scalebox{0.6}{\epsfig{file=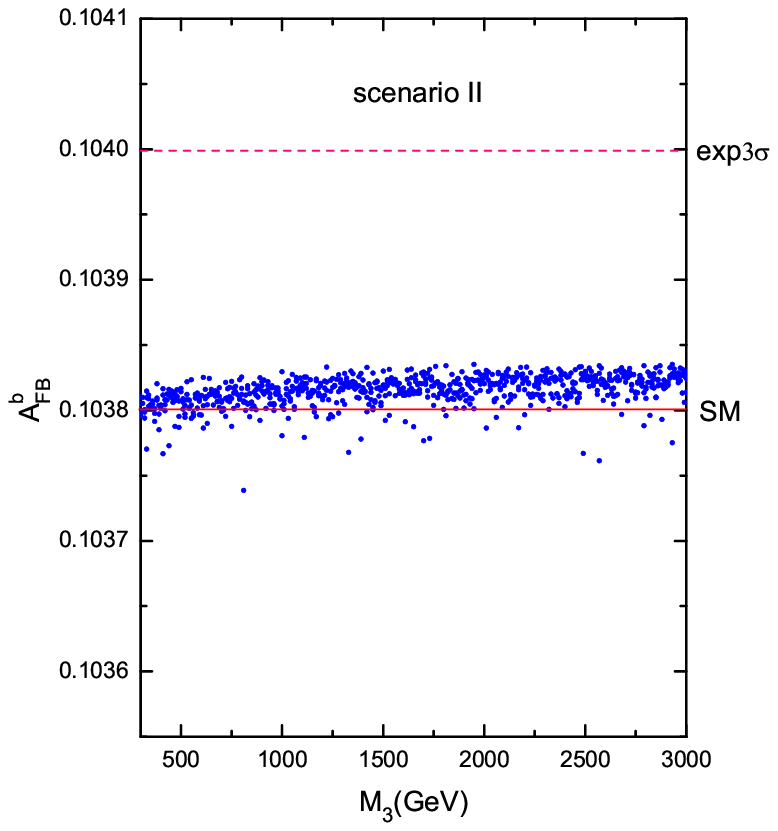}} \caption{Scatter plots of
$A_{FB}^{b}$ versus $M_{3}$ in three different scenarios,
respectively. The experimental value $A_{FB}^{b}=0.0992\pm0.0016$.}
\end{figure}

\begin{figure}
\scalebox{0.61}{\epsfig{file=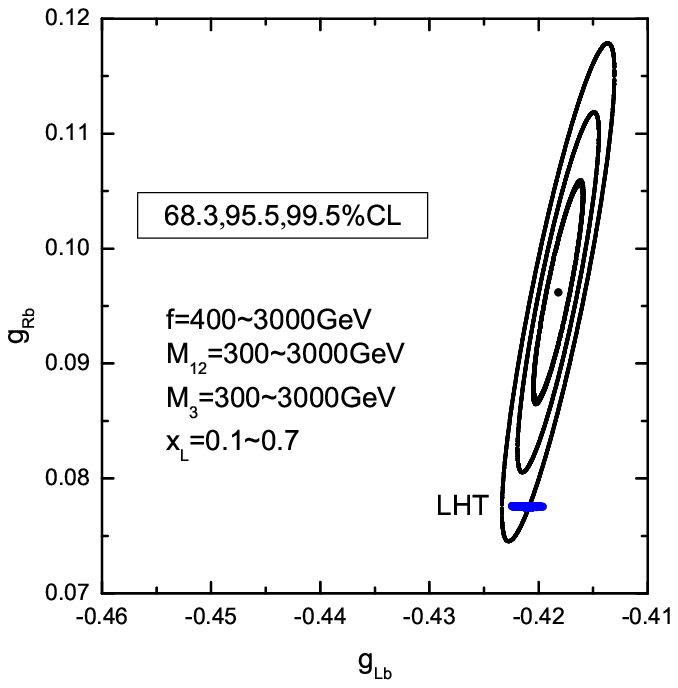}}
\scalebox{0.61}{\epsfig{file=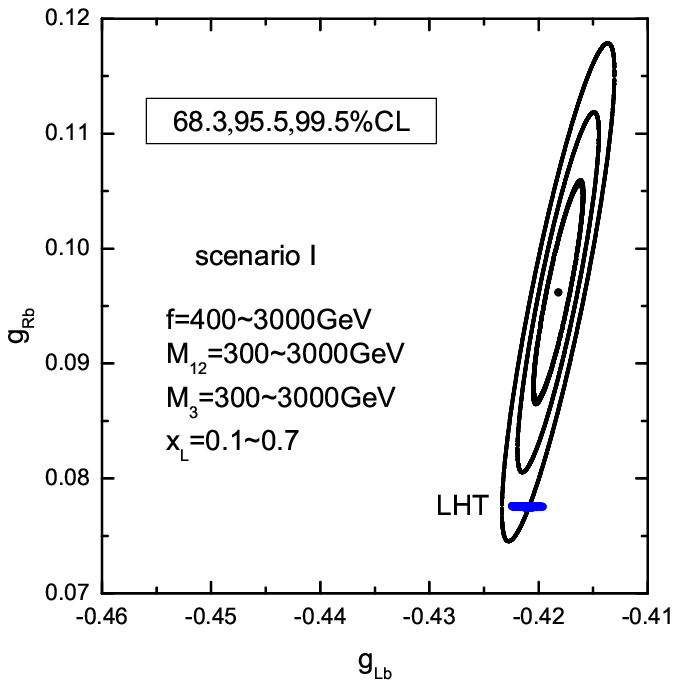}}
\scalebox{0.61}{\epsfig{file=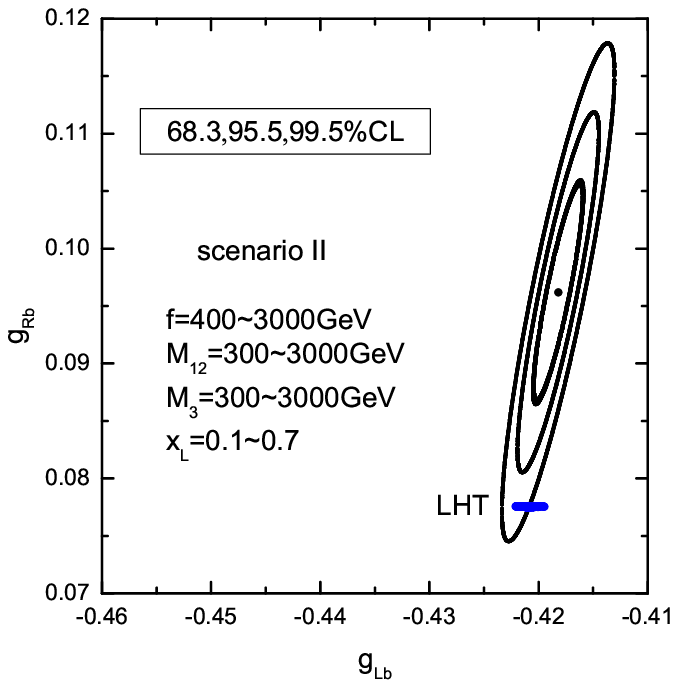}} \caption{The left-handed
and right-handed coupling constants in the LHT model. The
experimental value $g_{L}^{b}=-0.4182\pm0.0015,
g_{R}^{b}=0.0962\pm0.0063$.}
\end{figure}

\begin{figure}
\scalebox{0.61}{\epsfig{file=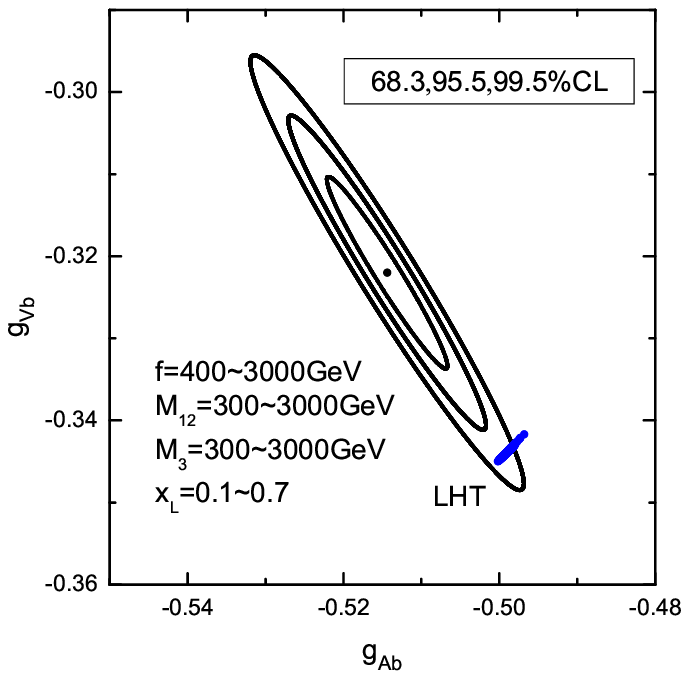}}
\scalebox{0.61}{\epsfig{file=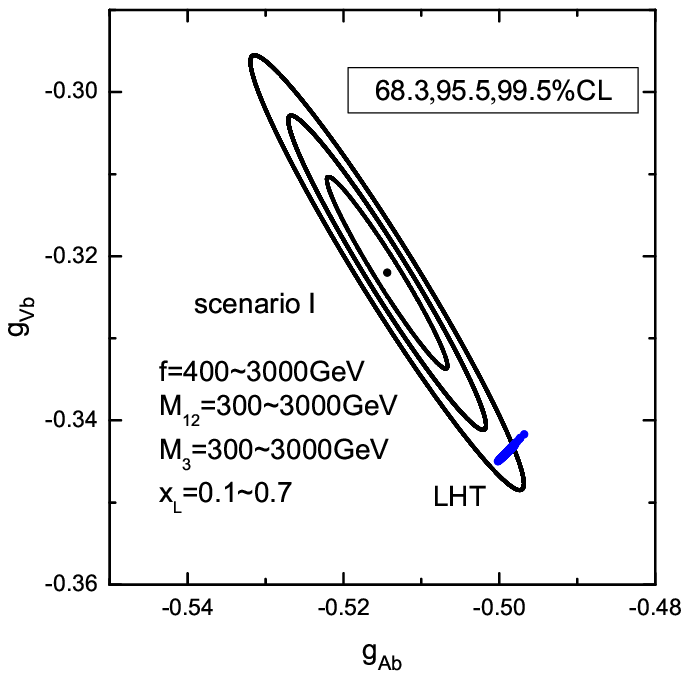}}
\scalebox{0.61}{\epsfig{file=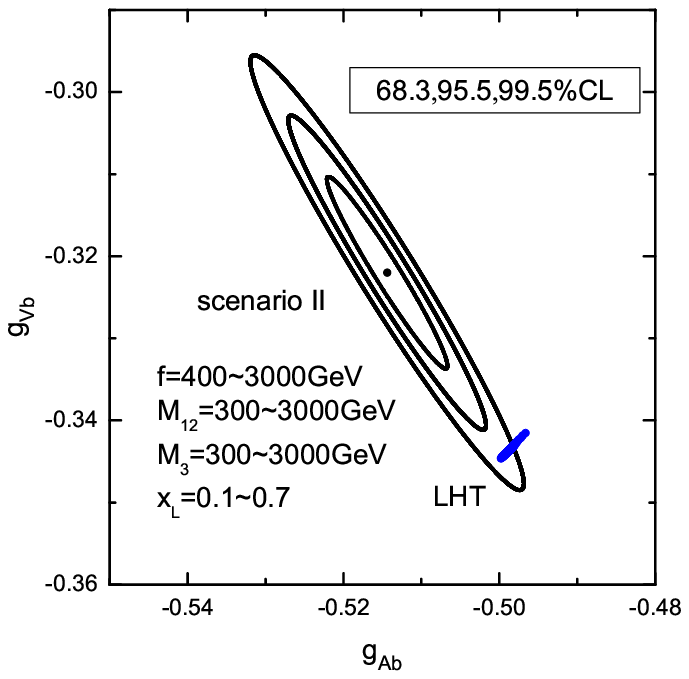}} \caption{The effective
vector and axial-vector coupling constants in the LHT model. The
experimental value $g_{V}^{b}=-0.3220\pm0.0077,
g_{A}^{b}=-0.5144\pm0.0051$.}
\end{figure}
Firstly, we discuss the $R_{b}$ changes with the LHT parameters, the
numerical results are summarized in Fig.(2-3). To see the influence
of the mixing parameter $x_{L}$ on the $R_{b}$, considering the
existing constraints, we let the parameters vary randomly in the
range: $M_{12}=300\sim3000GeV$, $M_{3}=300\sim3000GeV$,
$f=400\sim3000GeV$. In these three scenarios, we can see the plots
of $R_{b}$ decline with the $x_{L}$ increasing, which shows that the
contribution of the mixing diagrams between $t$ and $T^{+}$ is
negative and becomes larger with the $x_{L}$ increasing. When
$x_{L}>0.7$, part of the plots are beyond the $2\sigma$ regions of
its experimental value. This feature is similar in three different
scenarios.

To see the influence of the first two generation mirror quarks mass
$M_{12}$ on the $R_{b}$, considering the constraint from $R_{b}$ on
the $x_{L}$, we let the parameters vary randomly in the range:
$M_{3}=300\sim3000GeV$, $f=400\sim3000GeV$, $x_{L}=0.1\sim0.7$. In
these three scenarios, we can see the plots of $R_{b}$ are almost in
the $2\sigma$ regions of its experimental value. The noticeable
feature is that the $R_{b}$ isn't sensitive to $M_{12}$ so that the
constraint from $R_{b}$ on $M_{12}$ is very loose.

Secondly, we discuss the $A_{FB}^{b}$ changes with the LHT
parameters, the numerical results are summarized in Fig.(4-6). Same
as the $R_{b}$, the $A_{FB}^{b}$ isn't sensitive to $M_{12}$, so we
don't give the figures of the $A_{FB}^{b}$ as the function of
$M_{12}$. To see the influence of the mixing parameter $x_{L}$ on
the $A_{FB}^{b}$, we let the parameters vary randomly in the range:
$M_{12}=300\sim3000GeV$, $M_{3}=300\sim3000GeV$, $f=400\sim3000GeV$.
For the same reason, we can see the plots of $A_{FB}^{b}$ decline
and become closer to the experimental central value with the $x_{L}$
increasing. However, the contribution of the new particles is not
large enough so that the plots of the $A_{FB}^{b}$ are still
entirely scattered between the $2\sigma$ and $3\sigma$ region of its
experimental value.

To see the influence of the scale $f$ on the $A_{FB}^{b}$, we let
the parameters vary randomly in the range: $M_{12}=300\sim3000GeV$,
$M_{3}=300\sim3000GeV$, $x_{L}=0.1\sim0.7$. We can see the plots of
$A_{FB}^{b}$ are entirely between the $2\sigma$ and $3\sigma$ region
of its experimental value. The plots of $A_{FB}^{b}$ become closer
to the SM with the $f$ increasing, which shows that the contribution
of the heavy particles decouples with the $f$ increasing.

To see the influence of the third generation mirror quarks mass
$M_{3}$ on the $A_{FB}^{b}$, we let the parameters vary randomly in
the range: $M_{12}=300\sim3000GeV$, $f=400\sim3000GeV$,
$x_{L}=0.1\sim0.7$. We can see the plots of $A_{FB}^{b}$ are
entirely between the $2\sigma$ and $3\sigma$ region of its
experimental value.

Finally, we discuss the $Zbb$ couplings in the LHT model. In our
calculation, we still consider the above three scenarios and let the
parameters vary randomly in the range: $M_{12}=300\sim3000GeV$,
$M_{3}=300\sim3000GeV$, $f=400\sim3000GeV$, $x_{L}=0.1\sim0.7$, the
numerical results are summarized in Figs.(7-8). We confirm the
result of Ref.\cite{18}, in which the correction from the mixing
diagrams between $t$ and $T^{+}$ to $Zb\bar{b}$ couplings is mainly
on the $g_{L}^{b}$ and doesn't have the correct sign to alleviate
the large deviation between theoretical predictions and experimental
values. The plots scatter beyond the $3\sigma$ region their
experimental values are mainly caused by these couplings.
Furthermore, the correction on the $g_{R}^{b}$ is very small.
However, there is a little difference when we consider the
contributions involve other new particles. At this time, we can see
part of the plots scatter in the $3\sigma$ internal region of their
experimental values, where the deviation of $g_{L}^{b}$ can be
alleviated. Unfortunately, the correction on the $g_{R}^{b}$ is
still very small and the plots still scatter near the $3\sigma$
region of their experimental values so that the large deviation
between theoretical predictions and experimental values can't be
explained. The similar results are found on the $g_{A}^{b}$ and
$g_{V}^{b}$.

\section{Conclusions} \noindent In this paper,we studied the
one-loop contributions of the new particles to the $R_{b}$ and
$A_{FB}^{b}$ for three different scenarios in the framework of the
LHT model. From the scatter plots of $R_{b}$ versus $x_{L}$, the
precision measurement data of $R_{b}$ can give strong constraint on
the $x_{L}$. Considering this constraint, we can see $R_{b}$ isn't
sensitive to the mass of the first two generation mirror quarks. The
relevant parameters are weakly constrained by the precision
measurement data of $A_{FB}^{b}$. In the given parameters space, the
large deviation of $A_{FB}^{b}$ can't be explained reasonably. From
our study, the LHT model can provide the correction to the
$g_{L}^{b}$ and have small part of the parameter space to alleviate
the deviation between theoretical predictions and experimental
values. But the LHT model can't provide the large correction to the
$g_{R}^{b}$ so that the large deviation between the SM prediction
predictions and experimental values of the $Zbb$ couplings can't be
alleviated substantially.\vspace{4mm}
\\
\vspace{4mm} \textbf{Acknowledgments}\\
We would thank Junjie Cao and Lei Wu for useful discussions and
providing the calculation programs. This work is supported by the
National Natural Science Foundation of China under Grant
Nos.10775039, 11075045, by Specialized Research Fund for the
Doctoral Program of Higher Education under Grant No.20094104110001
and by HASTIT under Grant No.2009HASTIT004.

%\begin{appendix}
%\documentcalss[fleqn]{article}
\newpage
\begin{center}
\textbf{Appendix A: The expression of the renormalization vertex
$\hat{\Gamma}^{\mu}_{Zb\bar{b}}$} \cite{20}
\end{center}
\begin{figure}[th]
\scalebox{0.5}{\epsfig{file=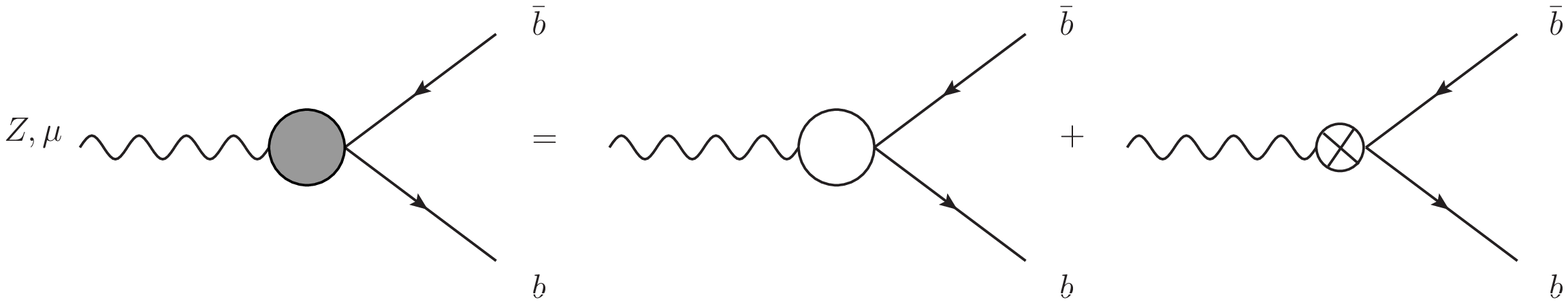}}
\end{figure}
\begin{eqnarray}
\hat{\Gamma}^{\mu}_{Zb\bar{b}}&=&\Gamma^{\mu}_{Zb\bar{b}}-ie\gamma^{\mu}(v_{b}-a_{b}\gamma_{5})\frac{C_{W}}{2S_{W}}
\delta Z_{ZA}-ieQ_{b}\gamma^{\mu}\frac{1}{2} \delta
Z_{ZA}\nonumber\\
&+&ie\gamma^{\mu}(v_{b}-a_{b}\gamma_{5})\delta
Z_{V}^{b}-ie\gamma^{\mu}\gamma_{5}(v_{b}-a_{b}\gamma_{5})\delta
Z_{A}^{b}\nonumber
\end{eqnarray}
where
\begin{eqnarray*}
v_{b}\equiv\frac{I_{b}^{3}-2Q_{b}S_{W}^{2}}{2C_{W}S_{W}},\quad
a_{b}\equiv\frac{I_{b}^{3}}{2C_{W}S_{W}},
\quad I_{b}^{3}=-\frac{1}{2},\quad Q_{b}=-\frac{1}{3}~~~~~~~~~~~\\
\quad \delta Z_{ZA}=2\frac{\Sigma_{T}^{AZ}(0)}{M_{Z_{L}}^{2}}~~~~~~~~~~~~~~~~~~~~~~~~~~~~~~~~~~~~~~~~~~~~~~~~~~~~~~~~~~~~~~~~~~\\
\delta
Z_{L}^{b}=Re\Sigma_{L}^{b}(m_{b}^{2})+m_{b}^{2}\frac{\partial}{\partial
P_{b}^{2}}Re[\Sigma_{L}^{b}(P_{b}^{2})+\Sigma_{R}^{b}(P_{b}^{2})+2\Sigma_{S}^{b}(P_{b}^{2})]|_{P_{b}^{2}=m_{b}^{2}}\\
\delta
Z_{R}^{b}=Re\Sigma_{R}^{b}(m_{b}^{2})+m_{b}^{2}\frac{\partial}{\partial
P_{b}^{2}}Re[\Sigma_{L}^{b}(P_{b}^{2})+\Sigma_{R}^{b}(P_{b}^{2})+2\Sigma_{S}^{b}(P_{b}^{2})]|_{P_{b}^{2}=m_{b}^{2}}\\
\delta Z_{V}^{b}=\frac{1}{2}(\delta Z_{L}^{b}+\delta
Z_{R}^{b}),\delta Z_{A}^{b}=\frac{1}{2}(\delta Z_{L}^{b}-\delta
Z_{R}^{b})~~~~~~~~~~~~~~~~~~~~~~~~~~~~~~~
\end{eqnarray*}
\begin{eqnarray*}
\hat{\Gamma}^{LHT,\mu}_{Zb\bar{b}}&=&\Gamma^{\mu}_{Zb\bar{b}}(\pi^{\pm})+\Gamma^{\mu}_{Zb\bar{b}}(\eta)+
\Gamma^{\mu}_{Zb\bar{b}}(\omega^{0})+\Gamma^{\mu}_{Zb\bar{b}}(\omega^{\pm})+
\Gamma^{\mu}_{Zb\bar{b}}(W_{L}^{\pm})+\Gamma^{\mu}_{Zb\bar{b}}(A_{H})+\Gamma^{\mu}_{Zb\bar{b}}(Z_{H})\\
&+&\Gamma^{\mu}_{Zb\bar{b}}(W_{H}^{\pm})+\Gamma^{\mu}_{Zb\bar{b}}(\pi^{\pm},W_{L}^{\pm})
+\Gamma^{\mu}_{Zb\bar{b}}(\omega^{\pm},W_{H}^{\pm})
+\delta\Gamma^{\mu}_{Zb\bar{b}}(\pi^{\pm})+\delta\Gamma^{\mu}_{Zb\bar{b}}(\eta)+
\delta\Gamma^{\mu}_{Zb\bar{b}}(\omega^{0})\\
&+&\delta\Gamma^{\mu}_{Zb\bar{b}}(\omega^{\pm})+\delta\Gamma^{\mu}_{Zb\bar{b}}(W_{L}^{\pm})
+\delta\Gamma^{\mu}_{Zb\bar{b}}(A_{H})+\delta\Gamma^{\mu}_{Zb\bar{b}}(Z_{H})+
\delta\Gamma^{\mu}_{Zb\bar{b}}(W_{H}^{\pm})
\end{eqnarray*}

\newpage
\begin{center}
\textbf{Appendix B: The explicit expressions of the $\delta
g_{L,R}^{LHT}$}
\end{center}
They can be represented in form of 1-point, 2-point and 3-point
standard functions $A,B_{0},B_{1},C_{ij}$. Here $P_{b}$ and
$\bar{P_{b}}$ are outgoing. In all expressions, the mass of b-quark
is ignored.
\begin{eqnarray*}
\delta
g_{L}&&=\frac{1}{16\pi^{2}}g^{2}C_{W}^{2}(V_{Hd})^{\ast}_{i3}(V_{Hd})_{i3}m_{u_{H}^{i}}^{2}
C_{0}^{a}\\
&&-\frac{1}{16\pi^{2}}\frac{g'^{2}}{100M_{A_{H}}^{2}}(V_{Hd})^{\ast}_{i3}(V_{Hd})_{i3}
\{(-\frac{1}{2}+\frac{1}{3}S_{W}^{2})[m_{d_{H}^{i}}^{4} C_{0}^{b}
-m_{d_{H}^{i}}^{2}M_{Z_{L}}^{2}C_{12}^{b}
-m_{d_{H}^{i}}^{2}M_{Z_{L}}^{2}C_{23}^{b}\\
&&-2m_{d_{H}^{i}}^{2}C_{24}^{b}+\frac{1}{2}m_{d_{H}^{i}}^{2}]-\frac{1}{2}[\frac{1}{2}m_{d_{H}^{i}}^{2}B_{0}
(-P_{b},m_{d_{H}^{i}},M_{A_{H}})\\
&&+\frac{1}{2}m_{d_{H}^{i}}^{2}(m_{d_{H}^{i}}^{2}-M_{A_{H}}^{2})\frac{\partial}{\partial
P_{b}^{2}}B_{0}
(-P_{b},m_{d_{H}^{i}},M_{A_{H}})]\\&&-\frac{1}{3}S_{W}^{2}[-\frac{1}{2}m_{d_{H}^{i}}^{2}B_{0}
(-P_{b},m_{d_{H}^{i}},M_{A_{H}})-\frac{1}{2}m_{d_{H}^{i}}^{2}(m_{d_{H}^{i}}^{2}-M_{A_{H}}^{2})\frac{\partial}{\partial
P_{b}^{2}}B_{0} (-P_{b},m_{d_{H}^{i}},M_{A_{H}})]\}\\
&&-\frac{1}{16\pi^{2}}\frac{g^{2}}{4M_{Z_{H}}^{2}}(V_{Hd})^{\ast}_{i3}(V_{Hd})_{i3}\{(-\frac{1}{2}+\frac{1}{3}S_{W}^{2})[m_{d_{H}^{i}}^{4}
C_{0}^{c}-m_{d_{H}^{i}}^{2}M_{Z_{L}}^{2}C_{12}^{c}-m_{d_{H}^{i}}^{2}M_{Z_{L}}^{2}C_{23}^{c}\\
&&-2m_{d_{H}^{i}}^{2}C_{24}^{c}+\frac{1}{2}m_{d_{H}^{i}}^{2}]-\frac{1}{2}[\frac{1}{2}m_{d_{H}^{i}}^{2}B_{0}
(-P_{b},m_{d_{H}^{i}},M_{Z_{H}})\\&&+\frac{1}{2}m_{d_{H}^{i}}^{2}(m_{d_{H}^{i}}^{2}-M_{Z_{H}}^{2})\frac{\partial}{\partial
P_{b}^{2}}B_{0}
(-P_{b},m_{d_{H}^{i}},M_{Z_{H}})]-\frac{1}{3}S_{W}^{2}[-\frac{1}{2}m_{d_{H}^{i}}^{2}B_{0}
(-P_{b},m_{d_{H}^{i}},M_{Z_{H}})\\&&-\frac{1}{2}m_{d_{H}^{i}}^{2}(m_{d_{H}^{i}}^{2}-M_{Z_{H}}^{2})\frac{\partial}{\partial
P_{b}^{2}}B_{0} (-P_{b},m_{d_{H}^{i}},M_{Z_{H}})]\}\\
&&-\frac{1}{16\pi^{2}}\frac{g^{2}}{2M_{W_{H}}^{2}}(V_{Hd})^{\ast}_{i3}(V_{Hd})_{i3}\{(\frac{1}{2}-\frac{2}{3}S_{W}^{2})[m_{u_{H}^{i}}^{4}
C_{0}^{d}-m_{u_{H}^{i}}^{2}M_{Z_{L}}^{2}C_{12}^{d}
-m_{u_{H}^{i}}^{2}M_{Z_{L}}^{2}C_{23}^{d}\\
&&-2m_{u_{H}^{i}}^{2}C_{24}^{d}+\frac{1}{2}m_{u_{H}^{i}}^{2}]-\frac{1}{2}[\frac{1}{2}m_{u_{H}^{i}}^{2}B_{0}
(-P_{b},m_{u_{H}^{i}},M_{W_{H}})\\&&+\frac{1}{2}m_{u_{H}^{i}}^{2}(m_{u_{H}^{i}}^{2}-M_{W_{H}}^{2})\frac{\partial}{\partial
P_{b}^{2}}B_{0}
(-P_{b},m_{u_{H}^{i}},M_{W_{H}})]-\frac{1}{3}S_{W}^{2}[-\frac{1}{2}m_{u_{H}^{i}}^{2}B_{0}
(-P_{b},m_{u_{H}^{i}},M_{W_{H}})\\&&-\frac{1}{2}m_{u_{H}^{i}}^{2}(m_{u_{H}^{i}}^{2}-M_{W_{H}}^{2})\frac{\partial}{\partial
P_{b}^{2}}B_{0}
(-P_{b},m_{u_{H}^{i}},M_{W_{H}})]+2C_{W}^{2}m_{u_{H}^{i}}^{2}C_{24}^{e}\}\\
&&-\frac{1}{16\pi^{2}}\frac{g'^{2}}{100}(V_{Hd})^{\ast}_{i3}(V_{Hd})_{i3}
\{(-\frac{1}{2}+\frac{1}{3}S_{W}^{2})[-2m_{d_{H}^{i}}^{2}C_{0}^{f}
+2M_{Z_{L}}^{2}C_{11}^{f} +2M_{Z_{L}}^{2}C_{23}^{f}\\
&&+4C_{24}^{f}-2]+[\frac{1}{2}B_{0}(-P_{b},m_{d_{H}^{i}},M_{A_{H}})+\frac{1}{2}(m_{d_{H}^{i}}^{2}-M_{A_{H}}^{2})\frac{\partial}{\partial
P_{b}^{2}}B_{0}(-P_{b},m_{d_{H}^{i}},M_{A_{H}})\\&&-\frac{1}{3}S_{W}^{2}B_{0}(-P_{b},m_{d_{H}^{i}},M_{A_{H}})-\frac{1}{3}
S_{W}^{2}(m_{d_{H}^{i}}^{2}-M_{A_{H}}^{2})\frac{\partial}{\partial
P_{b}^{2}}B_{0}
(-P_{b},m_{d_{H}^{i}},M_{A_{H}})]-\frac{1}{2}+\frac{1}{3}S_{W}^{2}\}
\end{eqnarray*}
\begin{eqnarray*}
&&-\frac{1}{16\pi^{2}}\frac{g^{2}}{4}(V_{Hd})^{\ast}_{i3}(V_{Hd})_{i3}
\{(-\frac{1}{2}+\frac{1}{3}S_{W}^{2})[-2m_{d_{H}^{i}}^{2}C_{0}^{g}
+2M_{Z_{L}}^{2}C_{11}^{g}
+2M_{Z_{L}}^{2}C_{23}^{g}\\
&&+4C_{24}^{g}-2]+[\frac{1}{2}B_{0}(-P_{b},m_{d_{H}^{i}},M_{Z_{H}})+\frac{1}{2}(m_{d_{H}^{i}}^{2}-M_{Z_{H}}^{2})\frac{\partial}{\partial
P_{b}^{2}}B_{0}(-P_{b},m_{d_{H}^{i}},M_{Z_{H}})\\&&-\frac{1}{3}S_{W}^{2}B_{0}(-P_{b},m_{d_{H}^{i}},M_{Z_{H}})-\frac{1}{3}
S_{W}^{2}(m_{d_{H}^{i}}^{2}-M_{Z_{H}}^{2})\frac{\partial}{\partial
P_{b}^{2}}B_{0} (-P_{b},m_{d_{H}^{i}},M_{Z_{H}})]-\frac{1}{2}+\frac{1}{3}S_{W}^{2}\}\\
&&+\frac{1}{16\pi^{2}}\frac{g^{2}}{2}(V_{Hd})^{\ast}_{i3}(V_{Hd})_{i3}
(\frac{1}{2}-\frac{2}{3}S_{W}^{2})[2m_{u_{H}^{i}}^{2}C_{0}^{h}
+2M_{Z_{L}}^{2}C_{11}^{h}+2M_{Z_{L}}^{2}C_{23}^{h}+4C_{24}^{h}-2]\\
&&+\frac{1}{16\pi^{2}}\frac{g^{2}}{2}(V_{Hd})^{\ast}_{i3}(V_{Hd})_{i3}
C_{W}^{2}[-2M_{Z_{L}}^{2}C_{0}^{i}
-2M_{Z_{L}}^{2}C_{11}^{i}-2M_{Z_{L}}^{2}C_{23}^{i}-12C_{24}^{i}+2]\\
&&+\frac{1}{16\pi^{2}}\frac{g^{2}}{2}(V_{Hd})^{\ast}_{i3}(V_{Hd})_{i3}[\frac{1}{2}B_{0}(-P_{b},m_{u_{H}^{i}},M_{W_{H}})
+\frac{1}{2}(m_{u_{H}^{i}}^{2}-m_{W_{H}}^{2})\frac{\partial}{\partial
P_{b}^{2}}B_{0}(-P_{b},m_{u_{H}^{i}},M_{W_{H}})\\&&-\frac{1}{3}S_{W}^{2}B_{0}(-P_{b},m_{u_{H}^{i}},M_{W_{H}})-\frac{1}{3}
S_{W}^{2}(m_{u_{H}^{i}}^{2}-M_{W_{H}}^{2})\frac{\partial}{\partial
P_{b}^{2}}B_{0} (-P_{b},m_{u_{H}^{i}},M_{W_{H}})-\frac{1}{2}+\frac{1}{3}S_{W}^{2}]\\
&&+\frac{1}{16\pi^{2}}\frac{g^{2}}{M_{Z_{L}}^{2}}C_{W}^{2}[-2A(M_{W_{H}})+2M_{W_{H}}^{2}B_{0}(0,M_{W_{H}},M_{W_{H}})+2M_{W_{H}}^{2}+M_{Z_{L}}^{2}B_{0}(0,M_{W_{H}},M_{W_{H}})]\\
&&-\frac{1}{16\pi^{2}}\frac{2g^{2}}{M_{Z_{L}}^{2}}\{\frac{2}{3}(\frac{1}{2}-\frac{2}{3}S_{W}^{2})[-\frac{2}{3}A(m_{u_{H}^{i}})
+\frac{2}{3}m_{u_{H}^{i}}^{2}B_{0}(0,m_{u_{H}^{i}},m_{u_{H}^{i}})+\frac{2}{3}m_{u_{H}^{i}}^{2}]\\
&&-\frac{1}{3}(-\frac{1}{2}+\frac{1}{3}S_{W}^{2})[-\frac{2}{3}A(m_{d_{H}^{i}})+\frac{2}{3}m_{d_{H}^{i}}^{2}B_{0}(0,m_{d_{H}^{i}},m_{d_{H}^{i}})
+\frac{2}{3}m_{d_{H}^{i}}^{2}]\\
&&-(-\frac{1}{2}+S_{W}^{2})[-\frac{2}{3}A(m_{l_{H}^{i}})+\frac{2}{3}m_{l_{H}^{i}}^{2}B_{0}(0,m_{l_{H}^{i}},m_{l_{H}^{i}})
+\frac{2}{3}m_{l_{H}^{i}}^{2}]\}\\
&&+\frac{g^2x^2_{L}}{4M^2_{W_{L}}}(1-2S_{W}^2)\frac{v^2}{f^2}(V_{CKM})^{\ast}_{tb}(V_{CKM})_{tb}\frac{1}{16\pi^{2}}[-2m^2_{T^{+}}C_{24}^{j}]\\
&&+\frac{g^2x^2_{L}}{2M^2_{W_{L}}}\frac{v^2}{f^2}(V_{CKM})^{\ast}_{tb}(V_{CKM})_{tb}\frac{1}{16\pi^{2}}[\frac{2}{3}S_{W}^2 m^4_{T^{+}}C_{0}^{k}-m^2_{T^{+}}M_{Z_{L}}^{2}C_{12}^{k}-\frac{2}{3}S_{W}^2 m^2_{T^{+}}C_{23}^{k}\\
&&-\frac{4}{3}S_{W}^2 C_{24}^{k}+\frac{1}{3}S_{W}^2 m^2_{T^{+}}]\\
&&+\frac{g^2x^2_{L}}{4M^2_{W_{L}}}\frac{v^2}{f^2}(V_{CKM})^{\ast}_{tb}(V_{CKM})_{tb}\frac{1}{16\pi^{2}}\{\frac{1}{2}m^2_TB_0(-P_b,m_{T^{+}},M_{W_{L}})\\
&&+\frac{1}{2}m^2_{T^{+}}(m^2_{T^{+}} -M^2_{W_L})\frac{\partial}{\partial P_{b}^{2}}B_0(-P_b,m_{T^{+}},M_{W_{L}})\\
&&-\frac{1}{3}S_{W}^2 [m^2_{T^{+}}B_0(-P_b,m_{T^{+}},M_{W_{L}})+ m^2_{T^{+}}(m^2_{T^{+}}-M^2_{W_L})\frac{\partial}{\partial P_{b}^{2}}B_0(-P_b,m_{T^{+}},M_{W_{L}})]\}\\
&&+\frac{g^2x^2_{L}}{4M^2_{W_{L}}}\frac{v^2}{f^2}(V_{CKM})^{\ast}_{tb}(V_{CKM})_{tb}\frac{1}{16\pi^{2}}[-m^2_{T^{+}}m^2_tC_{0}^{l}]\\
&&+\frac{g^2x^2_{L}}{4M^2_{W_{L}}}\frac{v^2}{f^2}(V_{CKM})^{\ast}_{tb}(V_{CKM})_{tb}\frac{1}{16\pi^{2}}[-m^2_{T^{+}}m^2_tC_{0}^{m}]\\
&&+\frac{g^2x^2_{L}}{2}C_{W}^2
\frac{v^2}{f^2}(V_{CKM})^{\ast}_{tb}(V_{CKM})_{tb}\frac{1}{16\pi^{2}}[-2M_{Z_{L}}^{2}C_{0}^{j}-2M_{Z_{L}}^{2}C_{11}^{j}-2M_{Z_{L}}^{2}C_{23}^{j}-12C_{24}^{j}+2]\\
&&+\frac{g^2x^2_{L}}{2}\frac{v^2}{f^2}(V_{CKM})^{\ast}_{tb}(V_{CKM})_{tb}\frac{1}{16\pi^{2}}\frac{2}{3}S_{W}^2 [m^2_{T^{+}}C_{0}^{k}-2M_{Z_{L}}^{2}C_{11}^{k}-2M_{Z_{L}}^{2}C_{23}^{k}-\frac{4}{3}C_{24}^{k}+2]\\
\end{eqnarray*}
\newpage
\begin{eqnarray*}
&&+\frac{g^2x^2_{L}}{4}\frac{v^2}{f^2}(V_{CKM})^{\ast}_{tb}(V_{CKM})_{tb}\frac{1}{16\pi^{2}}[2M_{Z_{L}}^{2}C_{11}^{l}+2M_{Z_{L}}^{2}C_{23}^{l}+4C_{24}^{l}-2]\\
&&+\frac{g^2x^2_{L}}{4}\frac{v^2}{f^2}(V_{CKM})^{\ast}_{tb}(V_{CKM})_{tb}\frac{1}{16\pi^{2}}[2M_{Z_{L}}^{2}C_{11}^{m}+2M_{Z_{L}}^{2}C_{23}^{m}+4C_{24}^{m}-2]\\
&&+\frac{g^2x^2_{L}}{2}\frac{v^2}{f^2}(V_{CKM})^{\ast}_{tb}(V_{CKM})_{tb}\frac{1}{16\pi^{2}}[\frac{1}{2}B_0(-P_b,m_{T^{+}},M_{W_L})\\
&&+(m^2_{T^{+}} -M^2_{W_L})\frac{\partial}{\partial P_{b}^{2}}B_0(-P_b,m_{T^{+}},M_{W_L})-\frac{1}{3}S_{W}^2 B_0(-P_b,m_{T^{+}},M_{W_L})\\
&&-\frac{1}{3}S_{W}^2 (m^2_{T^{+}}-M^2_{W_L})\frac{\partial}{\partial P_{b}^{2}}B_0(-P_b,m_{T^{+}},M_{W_L})-\frac{1}{2}+\frac{1}{3}S_{W}^2]\\
&&-\frac{g^2x^2_{L}}{2}\frac{v^2}{f^2}(V_{CKM})^{\ast}_{tb}(V_{CKM})_{tb}\frac{1}{16\pi^{2}}[\frac{2}{3}S_{W}^2 m^2_{T^{+}}C_{0}^{n}+2(1-\frac{2}{3}S_{W}^2) M_{Z_{L}}^{2}C_{11}^{n}\\
&&+2(1-\frac{2}{3}S_{W}^2) M_{Z_{L}}^{2}C_{23}^{n}+4(1-\frac{2}{3}S_{W}^2)C_{24}^{n}-2(1-\frac{2}{3}S_{W}^2)]\\
&&-\frac{g^2x^2_{L}}{2}C_{W}^2 \frac{v^2}{f^2}(V_{CKM})^{\ast}_{tb}(V_{CKM})_{tb}\frac{1}{16\pi^{2}}[-2M_{Z_{L}}^{2}C_{0}^{o}-2M_{Z_{L}}^{2}C_{11}^{o}-2M_{Z_{L}}^{2}C_{23}^{o}-12C_{24}^{o}+2]\\
&&-\frac{g^2x^2_{L}}{2}\frac{v^2}{f^2}(V_{CKM})^{\ast}_{tb}(V_{CKM})_{tb}\frac{1}{16\pi^{2}}[\frac{1}{2}B_0(-P_b,m_{t},M_{W_L})+\frac{1}{2}(m^2_t
-M^2_{W_L})\frac{\partial}{\partial
P_{b}^{2}}B_0(-P_b,m_{t},M_{W_L})\\
&&-\frac{1}{3}S_{W}^2 B_0(-P_b,m_{t},M_{W_L})
-\frac{1}{3}S_{W}^2 (m^2_t-M^2_{W_L})\frac{\partial}{\partial P_{b}^{2}}B_0(-P_b,m_{t},M_{W_L})-\frac{1}{2}+\frac{1}{3}S_{W}^2]\\
&&-\frac{g^2x^2_{L}}{2M^2_{W_{L}}}\frac{v^2}{f^2}(V_{CKM})^{\ast}_{tb}(V_{CKM})_{tb}\frac{1}{16\pi^{2}}[-m^4_t(1-\frac{2}{3}S_{W}^2)C_{0}^{o}-m^2_tM_{Z_{L}}^{2}C_{12}^{o}\\
&&-\frac{2}{3}S_{W}^2 m^2_tM_{Z_{L}}^{2}C_{23}^{o}-\frac{4}{3}m^2_tS_{W}^2 C_{24}^{o}(\Bar{P_b},P_b,m_{t},M_{W_{L}},m_{t})+\frac{1}{3}S_{W}^2 m^2_t]\\
&&-\frac{g^2x^2_{L}}{4M^2_{W_L}}(1-2S_{W}^2)\frac{v^2}{f^2}(V_{CKM})^{\ast}_{tb}(V_{CKM})_{tb}\frac{1}{16\pi^{2}}[-2m^2_tC_{24}^{o}]\\
&&-\frac{g^2x^2_{L}}{4M^2_{W_L}}\frac{v^2}{f^2}(V_{CKM})^{\ast}_{tb}(V_{CKM})_{tb}\frac{1}{16\pi^{2}}m^2_t\{\frac{1}{2}B_0(-P_b,m_{t},M_{W_{L}})\\
&&+\frac{1}{2}(m^2_t -M^2_{W_{L}})\frac{\partial}{\partial P_{b}^{2}}B_0(-P_b,m_{t},M_{W_{L}})\\
&&-\frac{1}{3}S_{W}^2[ B_0(-P_b,m_{t},M_{W_{L}})+(m^2_t-M^2_{W_{L}})\frac{\partial}{\partial P_{b}^{2}}B_0(-P_b,m_{t},M_{W_{L}})]\}\\
&&-\frac{g^2x^2_{L}}{2} \frac{v^2}{f^2}S_{W}^2(V_{CKM})^{\ast}_{tb}(V_{CKM})_{tb}\frac{1}{16\pi^{2}}[m^2_{T^{+}}C_{0}^{j}(\Bar{P_b},P_b,M_{W_{L}},m_{T^{+}},M_{W_L})]\\
&&-\frac{g^2x^2_{L}}{2} \frac{v^2}{f^2}S_{W}^2(V_{CKM})^{\ast}_{tb}(V_{CKM})_{tb}\frac{1}{16\pi^{2}}[m^2_{T^{+}}C_{0}^{j}(\Bar{P_b},P_b,M_{W_L},m_{T^{+}},M_{W_{L}})]\\
&&+\frac{g^2x^2_{L}}{2} \frac{v^2}{f^2}S_{W}^2(V_{CKM})^{\ast}_{tb}(V_{CKM})_{tb}\frac{1}{16\pi^{2}}[m^2_tC_{0}^{o}(\Bar{P_b},P_b,M_{W_{L}},m_{t},M_{W_L})]\\
&&+\frac{g^2x^2_{L}}{2} \frac{v^2}{f^2}S_{W}^2(V_{CKM})^{\ast}_{tb}(V_{CKM})_{tb}\frac{1}{16\pi^{2}}[m^2_tC_{0}^{o}(\Bar{P_b},P_b,M_{W_L},m_{t},M_{W_{L}})]\\
\end{eqnarray*}
\begin{eqnarray*}
C_{ij}^{a}&=&C_{ij}^{a}(\bar{P_{b}},P_{b},M_{W_{H}},m_{u_{H}^{i}},M_{W_{H}})\\
C_{ij}^{b}&=&C_{ij}^{b}(\bar{P_{b}},P_{b},m_{d_{H}^{i}},M_{A_{H}},m_{d_{H}^{i}})\\
C_{ij}^{c}&=&C_{ij}^{c}(\bar{P_{b}},P_{b},m_{d_{H}^{i}},M_{Z_{H}},m_{d_{H}^{i}})\\
C_{ij}^{d}&=&C_{ij}^{d}(\bar{P_{b}},P_{b},m_{u_{H}^{i}},M_{W_{H}},m_{u_{H}^{i}})\\
C_{ij}^{e}&=&C_{ij}^{e}(\bar{P_{b}},P_{b},M_{W_{H}},m_{u_{H}^{i}},M_{W_{H}})\\
C_{ij}^{f}&=&C_{ij}^{f}(\bar{P_{b}},P_{b},m_{d_{H}^{i}},M_{A_{H}},m_{d_{H}^{i}})\\
C_{ij}^{g}&=&C_{ij}^{g}(\bar{P_{b}},P_{b},m_{d_{H}^{i}},M_{Z_{H}},m_{d_{H}^{i}})\\
C_{ij}^{h}&=&C_{ij}^{h}(\bar{P_{b}},P_{b},m_{u_{H}^{i}},M_{W_{H}},m_{u_{H}^{i}})\\
C_{ij}^{i}&=&C_{ij}^{i}(\bar{P_{b}},P_{b},M_{W_{H}},m_{u_{H}^{i}},M_{W_{H}})\\
C_{ij}^{j}&=&C_{ij}^{j}(\bar{P_b},P_b,M_{W_{L}},m_{T^{+}},M_{W_{L}})\\
C_{ij}^{k}&=&C_{ij}^{k}(\Bar{P_b},P_b,m_{T^{+}},M_{W_{L}},m_{T^{+}})\\
C_{ij}^{l}&=&C_{ij}^{l}(\Bar{P_b},P_b,m_{t},M_{W_{L}},m_{T^{+}})\\
C_{ij}^{m}&=&C_{ij}^{m}(\Bar{P_b},P_b,m_{T^{+}},M_{W_L},m_{t})\\
C_{ij}^{n}&=&C_{ij}^{n}(\Bar{P_b},P_b,m_{t},M_{W_L},m_{t})\\
C_{ij}^{o}&=&C_{ij}^{o}(\Bar{P_b},P_b,M_{W_L},m_{t},M_{W_L})~~~~~~~~~~~~~~~~~~~~~~~~~~~~~~~~~~~~~~~~~~~~~~~~~~~~~~~~~~~~~~~~~
\end{eqnarray*}
\begin{eqnarray*}
\delta
g_{R}&=&-\frac{1}{16\pi^{2}}\frac{g'^{2}}{100M_{A_{H}}^{2}}(V_{Hd})^{\ast}_{i3}(V_{Hd})_{i3}\{\frac{1}{3}S_{W}^{2}m_{d_{H}^{i}}^{2}B_{1}
(-P_{b},m_{d_{H}^{i}},M_{A_{H}})\\&&-\frac{1}{3}S_{W}^{2}[-\frac{1}{2}m_{d_{H}^{i}}^{2}B_{0}
(-P_{b},m_{d_{H}^{i}},M_{A_{H}})
-\frac{1}{2}m_{d_{H}^{i}}^{2}(m_{d_{H}^{i}}^{2}-M_{A_{H}}^{2})\frac{\partial}{\partial
P_{b}^{2}}B_{0}(-P_{b},m_{d_{H}^{i}},M_{A_{H}})]\}\\
&&-\frac{1}{16\pi^{2}}\frac{g^{2}}{4M_{Z_{H}}^{2}}(V_{Hd})^{\ast}_{i3}(V_{Hd})_{i3}\{\frac{1}{3}S_{W}^{2}m_{d_{H}^{i}}B_{1}
(-P_{b},m_{d_{H}^{i}},M_{Z_{H}})\\&&-\frac{1}{3}S_{W}^{2}[-\frac{1}{2}m_{d_{H}^{i}}^{2}B_{0}
(-P_{b},m_{d_{H}^{i}},M_{Z_{H}})
-\frac{1}{2}m_{d_{H}^{i}}^{2}(m_{d_{H}^{i}}^{2}-M_{Z_{H}}^{2})\frac{\partial}{\partial
P_{b}^{2}}B_{0}(-P_{b},m_{d_{H}^{i}},M_{Z_{H}})]\}\\
&&+\frac{1}{16\pi^{2}}\frac{g^{2}}{2M_{W_{H}}^{2}}(V_{Hd})^{\ast}_{i3}(V_{Hd})_{i3}\{\frac{1}{3}S_{W}^{2}m_{u_{H}^{i}}B_{1}
(-P_{b},m_{u_{H}^{i}},M_{W_{H}})\\&&-\frac{1}{3}S_{W}^{2}[-\frac{1}{2}m_{u_{H}^{i}}^{2}B_{0}
(-P_{b},m_{u_{H}^{i}},M_{W_{H}})
-\frac{1}{2}m_{u_{H}^{i}}^{2}(m_{u_{H}^{i}}^{2}-M_{W_{H}}^{2})\frac{\partial}{\partial
P_{b}^{2}}B_{0}(-P_{b},m_{u_{H}^{i}},M_{W_{H}})]\}\\
&&-\frac{1}{16\pi^{2}}\frac{g'^{2}}{100}(V_{Hd})^{\ast}_{i3}(V_{Hd})_{i3}\{\frac{2}{3}S_{W}^{2}B_{1}
(-P_{b},m_{d_{H}^{i}},M_{A_{H}})+\frac{1}{3}S_{W}^{2}B_{0}(-P_{b},m_{d_{H}^{i}},M_{A_{H}})\\
&&+\frac{1}{3}S_{W}^{2}(m_{d_{H}^{i}}^{2}-M_{A_{H}}^{2})\frac{\partial}{\partial
P_{b}^{2}}B_{0}(-P_{b},m_{d_{H}^{i}},M_{A_{H}})\}\\
&&-\frac{1}{16\pi^{2}}\frac{g^{2}}{4}(V_{Hd})^{\ast}_{i3}(V_{Hd})_{i3}\{\frac{2}{3}S_{W}^{2}B_{1}
(-P_{b},m_{d_{H}^{i}},M_{Z_{H}})+\frac{1}{3}S_{W}^{2}B_{0}(-P_{b},m_{d_{H}^{i}},M_{Z_{H}})\\
&&+\frac{1}{3}S_{W}^{2}(m_{d_{H}^{i}}^{2}-M_{Z_{H}}^{2})\frac{\partial}{\partial
P_{b}^{2}}B_{0}(-P_{b},m_{d_{H}^{i}},M_{Z_{H}})\}\\
&&-\frac{1}{16\pi^{2}}\frac{g^{2}}{2}(V_{Hd})^{\ast}_{i3}(V_{Hd})_{i3}\{\frac{2}{3}S_{W}^{2}B_{1}
(-P_{b},m_{u_{H}^{i}},M_{W_{H}})+\frac{1}{3}S_{W}^{2}B_{0}(-P_{b},m_{u_{H}^{i}},M_{W_{H}})\\
&&+\frac{1}{3}S_{W}^{2}(m_{u_{H}^{i}}^{2}-M_{W_{H}}^{2})\frac{\partial}{\partial
P_{b}^{2}}B_{0}(-P_{b},m_{u_{H}^{i}},M_{W_{H}})\}\\
&&+\frac{g^2x^2_{L}}{4M^2_{W_{L}}}\frac{v^2}{f^2}(V_{CKM})^{\ast}_{tb}(V_{CKM})_{tb}\frac{1}{16\pi^{2}}\{-\frac{2}{3}S_{W}^2 m^2_{T^{+}}B_{1}(-P_b,m_{T^{+}},M_{W_{L}})\\
&&-\frac{2}{3}S_{W}^2[\frac{1}{2}m^2_{T^{+}}B_{0}(-P_b,m_{T^{+}},M_{W_{L}})+\frac{1}{2}m^2_{T^{+}}(m^2_{T^{+}}-M^2_{W_L})\frac{\partial}{\partial P_{b}^{2}}B_0(-P_b,m_{T^{+}},M_{W_{L}})]\}\\
&&+\frac{g^2x^2_{L}}{2}\frac{v^2}{f^2}(V_{CKM})^{\ast}_{tb}(V_{CKM})_{tb}\frac{1}{16\pi^{2}}\{-\frac{2}{3}S_{W}^2 B_{1}(-P_b,m_{T^{+}},M_{W_{L}})\\
&&-\frac{1}{3}S_{W}^2[B_{0}(-P_b,m_{T^{+}},M_{W_{L}})+(m^2_{T^{+}}-M^2_{W_L})\frac{\partial}{\partial P_{b}^{2}}B_0(-P_b,m_{T^{+}},M_{W_{L}})]\}\\
&&-\frac{g^2x^2_{L}}{4M^2_{W_{L}}}\frac{v^2}{f^2}(V_{CKM})^{\ast}_{tb}(V_{CKM})_{tb}\frac{1}{16\pi^{2}}\{-\frac{2}{3}S_{W}^2 m^2_tB_{1}(-P_b,m_{t},M_{W_{L}})\\
&&-\frac{1}{3}S_{W}^2[m^2_tB_{0}(-P_b,m_{t},M_{W_{L}})+m^2_t(m^2_t-M^2_{W_L})\frac{\partial}{\partial P_{b}^{2}}B_0(-P_b,m_{t},M_{W_{L}})]\}\\
&&-\frac{g^2x^2_{L}}{2}\frac{v^2}{f^2}(V_{CKM})^{\ast}_{tb}(V_{CKM})_{tb}\frac{1}{16\pi^{2}}\{-\frac{2}{3}S_{W}^2 B_{1}(-P_b,m_{t},M_{W_{L}})\\
&&-\frac{1}{3}S_{W}^2[B_{0}(-P_b,m_{t},M_{W_{L}})+(m^2_t-M^2_{W_L})\frac{\partial}{\partial P_{b}^{2}}B_0(-P_b,m_{t},M_{W_{L}})]\}\\
\end{eqnarray*}

%\end{appendix}
%\newpage

\end{document}